\documentclass[fleqn,usenatbib]{mnras}

\usepackage{newtxtext,newtxmath}

\usepackage[T1]{fontenc}

\DeclareRobustCommand{\VAN}[3]{#2}
\let\VANthebibliography\thebibliography
\def\thebibliography{\DeclareRobustCommand{\VAN}[3]{##3}\VANthebibliography}

\usepackage{graphicx}	
\usepackage{amsmath}	
\usepackage[T1]{fontenc}
\usepackage{ae,aecompl}
\usepackage{longtable}
\usepackage{float}
\usepackage{bm}
\usepackage{pifont}
\usepackage{caption}
\usepackage{geometry}
\usepackage{tikz}
\usetikzlibrary{shapes.geometric, arrows}
\usepackage{adjustbox}
\usepackage{balance}
\usepackage{orcidlink}

\usepackage{array}

\title[sGRBs alike GW/GRB 170817]{Similar Fermi-GBM sGRBs to GW/sGRB 170817A in MeV-GeV energies}

\author[S. Rao \& Moharana]{
Sanjeeva Rao Prattipati,$^{1}$\orcidlink{0009-0004-9809-8372}\thanks{E-mail:p23ph0011@iitj.ac.in}
Reetanjali Moharana$^{{1}}$\orcidlink{ 0000-0001-5642-8311}
 \thanks {reetanjali@iitj.ac.in} and Sourav Dutta
\\
$^{1}$Department of Physics, Indian Institute of Technology Jodhpur,  
  Jodhpur-342037, India.}

\date{Accepted XXX. Received YYY; in original form ZZZ}

\pubyear{\the\year{2025}}

\begin{document}
\label{firstpage}
\pagerange{\pageref{firstpage}--\pageref{lastpage}}
\maketitle

\begin{abstract}
The rate of observed gravitational waves (GWs) from neutron star-neutron star (NS-NS) mergers detected by the Laser Interferometer Gravitational-Wave Observatory (LIGO) indicates the existence of more than one short gamma-ray bursts (sGRBs) similar to GW/sGRB 170817A within the total gamma-ray bursts (GRBs) recorded by satellite detectors such as BATSE, Fermi-Gamma-ray Burst Monitor (Fermi-GBM), and Swift-Burst Alert Telescope (Swift-BAT). We investigated sGRBs in the Fermi-GBM dataset based on their MeV-GeV $\gamma$-ray emission features, to identify sGRBs similar to sGRB 170817A. Any addition of such events can impact the rate of NS-NS CBC events observed by LIGO. SGRB 170817A exhibits two distinct emission components: a non-thermal peak and a thermal component. We adopted a multifaceted approach to identify analogous sGRBs, which involved computing the hardness ratios $HR_{1}$ and $HR_{2}$ and then clustering them via the K-means algorithm. Our further studies reveal the presence of eight such events in Fermi-GBM data, which will enable us to calculate the rate of electromagnetic (EM) counterparts associated with LIGO GW events (GW+sGRB events) across all observing runs. Giving an estimation, by the end of the $O_4$ LIGO run, there could be nearly 5 GW+sGRB events. Deviation from this number may raise concerns about our understanding of the evolution of such events over distance. 
\end{abstract}




\section{Introduction}\label{intro}

The Advanced LIGO and Virgo gravitational wave (GW) radiation observatories detected the only confirmed binary neutron star (BNS) merger on August 17, 2017, through multi-wavelength observations. The event was subsequently identified as sGRB 170817A, which triggered 1.7 seconds after the GW event. The MeV-GeV emissions observed by Fermi-GBM report the event with T$_{90}${\footnote{$T_{90}$ is the duration in which the central 90\% of the gamma-ray signal is detected.}} = $2.048$ sec \citep{2017ApJ...848L..14G}. The micronovae/kilonovae observed around the event in optical,  UV, and IR as a counterpart by SSS17a \citep{Coulter:2017wya,2017ApJ...851L..21V,Drout:2017ijr,Evans:2017mmy,2017ApJ...848L..27T,Smartt:2017fuw,Pian:2017gtc} confirms the presence of heavy elements originated by the r-process of the NS-NS merger. This multimessenger observation confirmed the long-standing theory that one of the origins of sGRBs is the merger of compact objects, such as neutron stars \citep{poggiani2020multi,Eichler:1989ve,Narayan:1992iy}. 
Previously, a pulsar survey on NS-NS objects estimation in the near sky (both Galactic and Extragalactic) predicted a reasonable number of BNS events to be observed by LIGO \citep{Kalogera:2001dz}, where the original idea is given in \citep{Narayan:1991fn, Kalogera:2003tn}. Reference  \citep{Kalogera:2001dz} predicted the NS-NS inspiral detection rates with the then proposed LIGO II, a rate of 2-1300 yr $^{-1}$ within a distance of 350 Mpc, comparable to the upcoming $O_5$ run of LIGO. Theoretically, the other way to estimate the rate of NS-NS mergers is through population synthesis \citep{2007ApJ...662..504B}. A phenomenological approach based on the rate of sGRBs observed by Swift, with the well-known redshift \citep{2012MNRAS.425.2668C}, leads to predictions of lower and upper detection rate limits of (1-180) yr$^{-1}$ by Advanced LIGO (aLIGO). The rate density of sGRBs used for this calculation is $8^{+5}_{-3} -$
 $1100^{+700}_{-470}\ \mathrm{Gpc^{-3}\ yr^{-1}}$.  

Interestingly, later observations with Very Long Baseline Interferometry (VLBI) of the sGRB 170817A revealed a super-luminal motion of the later radio afterglow image \citep{Mooley:2018qfh}. These observations imply that a powerful relativistic jet broke out from the ejecta 
making an angle of $20 \pm 5$ degrees off to us. The observation suggests a jet opening angle $\theta_{j} \le 5$ deg and a difference with the viewing angle of $(\theta_{obs} - \theta_{j}) \sim 0.2-0.35$ rad. Prior to this event, the existence of a GRB originating from such a highly off-axis jet was uncertain. Although models suggest sGRB viewing angles can range from $18^\circ$ to $33^\circ$  \citep{Granot:2017gwa,Howell:2018nhu}, this implies that systems with inclinations above $33^\circ$ or below $147^\circ$ may not produce observable GRB emission. Additionally, the work in \citep{Mazwi:2024ego} using BNS events from LIGO (GW170817 and GW190425), Neutron star-Black hole (NS-BH) (GW190917\_114630 and GW200115\_042309), and GW190425 suggests that the reason for the further absence of compact binary coalescence (CBC) with MeV-GeV emissions is due to the large inclinations. Hence, a detailed study on the visibility of sGRBs at off-axis angles, both theoretically and through significant observations, can also impact the study of CBC GW rates with EM counterparts.

According to further relevant works where only the LIGO detected BNS mergers events have been considered, the event rate of GW events with EM counterparts is calculated as (GW/sGRBs 170817A), R  = $1540_{-1220}^{+3200}$ Gpc$^{-3}$ yr$^{-1}$ \citep{LIGOScientific:2017vwq,KAGRA:2021duu},     
While recent findings, utilizing updated LIGO sensitivity and detection volume, yield a new rate estimated as $680-1300$ Gpc$^{-3}$ yr$^{-1}$ \citep{Hayes:2023zxm}.
This rate implies that BNS mergers are rather frequent and play a significant role in the universe's r-process nucleosynthesis. Additional detections with the ongoing GW detectors network, including LIGO, Virgo, and the Japanese Kamioka Gravitational Wave Detector (KAGRA), along with the upcoming LIGO India \citep{Pandey:2024mlo}, aim to provide a precise measurement of the rate estimate of BNS events by looking deeper into the sky. In the meantime, an estimation of the rate by examining the sGRB behaviors of such events can provide a better understanding to tune the observations. Hence, a detailed statistical interpretation is necessary to identify similar sGRBs to sGRB 170817A among the previously observed sGRBs by the three gamma-ray satellite detectors: BATSE, Fermi-GBM, and Swift. A similar attempt has been made by several groups, including the Fermi-GBM collaboration \citep{2019ApJ...876...89V} and the Swift team, which detected sGRBs \citep{Kapadia:2024elj}. In this work, we present a statistically detailed study of the Fermi-GBM sGRBs, complementing and adding new results to the aforementioned studies.  

The peculiarity of the sGRB 170817A follows the presence of two distinct emission components over the T$_{90}$ of 2.048 sec. The first emission from -0.192 sec to 1.216 sec is a soft pulse with alpha ($\alpha$) = $0.85 \pm 1.38$ and peak emission E$_{\text{peak}}$ = $215 \pm 54$ keV, and the second emission, within 1.216 sec to 1.856 sec follows a softer thermal tail at the energy kT $ = 10.3 \pm 1.5$ keV \citep{2017ApJ...848L..14G,2019ApJ...876...89V}. This peculiar feature is explained as being observed with an off-axis viewing angle of the jet \citep{Troja:2017nqp,2017ApJ...848L..20M,Fong:2017ekk,2017ApJ...850L..19M}. Our investigation focuses on identifying and characterizing dual-peak structures similar to those observed in other sGRB light curves, with an emphasis on distinguishing between thermal and nonthermal features. Although clearly the unique feature of the sGRB is its low luminosity, $L_{iso}$ = $1.6 \times 10^{47}$ erg $s^{-1}$ \citep{Zhang:2017lpb}. Unfortunately, without knowing the distances of each sGRB, we cannot make this a search criterion. 

The structure of the paper is as follows: Section \ref{rate} explains the calculation of the expected number of sGRBs for all major satellite detectors up to 17 August 2017, following the LIGO observed rate for BNS. Section \ref{archive} describes the selection of the sGRB events from the Fermi-GBM catalog for their {HR\textsubscript{1} and HR\textsubscript{2} calculation. The detailed tools used for analysis are explained in section \ref{tool}. In subsection \ref{ml} we discuss the machine learning (ML) model K-means clustering for the classification of sGRBs, and further in subsection \ref{LC-SED} we discuss the light curve and spectral analysis. Finally, in section \ref{result} and section \ref{discussion}, we present the obtained results, a summary, and the future scope of this work.

\section{Expected \MakeLowercase{s}grb rate}\label{rate}

The EM counterpart of a GW event was observed and identified as the short GRB 170817A, which originated from an NS-NS merger, during the second Observing run ($O_2$) of Advanced LIGO. The running time of the $O_2$ run is from
November 30, 2016, to August 25, 2017. Remarkably, the rate of such an observation recorded as R = $1540_{-1220}^{+3200}$ Gpc$^{-3}$ yr$^{-1}$ \citep{LIGOScientific:2017vwq}, which was quite high compared to the expected rate of NS-NS merger from theoretical estimations. A detailed further analysis and interpretation predict the rate of such events with LIGO to be $R_{GW}= 680-1300$ Gpc$^{-3}$ yr$^{-1}$ \citep{Hayes:2023zxm}. Considering both these rates of sGRBs, we can calculate the total number of such events that must have been listed in the major satellite detectors.
\begin{equation}
\label{eq:rate}
 N_{obs} = R_{GW} \, V_{x}  \,  \mathbb{T} \, \mathbb{D}  \left(\frac{\Delta\Omega_{x} }{4\pi}\right).
 \end{equation} 
$r_x$ is the distance till which a sGRB mimicking sGRB170817A is significantly above the background. Reference \citep{Zhang:2017lpb} showed that the first peak would be visible if the GRB is at a distance of a maximum of 65 Mpc, whereas the second peak would merge with the background at a distance of 55 Mpc. Hence we have taken $r_x = 65$ Mpc and 55 Mpc, and the corresponding volume, $V_x = \left(\frac{4}{3}\right) \, \pi \, r_x^3$. $\mathbb{D}$ is the duty cycle of the detector, and $\mathbb{T}$ is the corresponding detectors total operating time. The values of these parameters for the major detectors, Fermi-GBM, BATSE, and Swift, are listed in Table \ref{table_number}. The details of the expected sGRBs within the detector data are also listed in Table \ref{table_number}. It is worth noting that, based on the Fermi-GBM data, a total of 5.06 sGRBs are expected to originate from BNS systems at a distance of 65 Mpc, assuming a rate of 1300 Gpc$^{-3}$ yr$^{-1}$. Note that the total number calculation excludes the estimation of the off-axis angle, which may further reduce the estimated number. 

\begin{table*}
\fontsize{6}{6}
	\centering
	\caption{Number of events for different detectors following the LIGO rate for NS-NS mergers.
* denotes that  the partial  sky coverage of BATSE is included in its  effective duty cycle, $\mathbb{D}$ is taken from these ref, for Fermi-GBM  :\citep{Meegan:1998tn}, BATSE \& Swift:\citep{Porciani:2000ag}}
	\label{table_number}
	\begin{tabular}{lccccccccccr} 
		\hline
		Detectors &$\mathbb{T}$& $\boldsymbol{\Delta\Omega}$ [sr] & $\mathbb{D}$ &  \multicolumn{6}{c}{Number of events} \\
		\hline
        & &                                 & & \multicolumn{1}{c}{65 Mpc}& \multicolumn{1}{c}{55 Mpc} & \multicolumn{2}{c}{65 Mpc} & \multicolumn{2}{c}{55 Mpc} \\
         \cline{5-10}
         & &                                 & & $\begin{array}{c}
1540 \\
\text{Gpc}^{-3}\text{yr}^{-1}
\end{array}$ & $\begin{array}{c}
1540 \\
\text{Gpc}^{-3}\text{yr}^{-1}
\end{array}$  & $\begin{array}{c}
1300 \\
\text{Gpc}^{-3}\text{yr}^{-1}
\end{array}$   & $\begin{array}{c}
680 \\
\text{Gpc}^{-3}\text{yr}^{-1}
\end{array}$
  & $\begin{array}{c}
1300 \\
\text{Gpc}^{-3}\text{yr}^{-1}
\end{array}$
   & $\begin{array}{c}
680 \\
\text{Gpc}^{-3}\text{yr}^{-1}
\end{array}$
    \\
 \hline
Fermi-GBM & 17 Jul, 2008 - 17 Aug, 2017 &9.5 & 0.5 &6.08 &3.68 &5.06 &2.65& 3.07&1.60\\
BATSE &21 Apr, 1991 - 17 Aug, 2000 &4$\pi ^*$ & 0.456 &7.54 & 4.57& 6.28&3.28 & 3.80&1.99\\
Swift  &17 Dec, 2004 - 17 Aug, 2017& 1.4 & 0.78 &1.95& 1.18 & 1.62 &0.84& 0.98&0.51\\
		\hline
	\end{tabular}
\end{table*}

\section{Methodology for selection of \MakeLowercase{s}GRB events}\label{archive}

For simplicity and clarity, we restrict our analysis to only one detector, Fermi-GBM, for our subsequent statistical studies. In a future study, we would expand the same analysis, including all the three major detectors, BATSE, Swift, and Fermi-GBM, where we will have to consider calibration between the detectors. After the launch of the Gamma-ray Large Area Space Telescope (GLAST), Fermi-GBM began detecting GRBs on July 17, 2008. To date, it has observed more than 4091 GRBs {\footnote{https://heasarc.gsfc.nasa.gov/W3Browse/fermi/fermigbrst.html}} with a rate of nearly one GRB per day. The detector covers almost the full sky in one orbit. 

We collected the sGRBs from the Fermi-GBM catalog \citep{Bhat:2016odd} available on the NASA HEASARC website. Subsequently following steps are adopted to form the sample twining of sGRB 170817A. 

\begin{enumerate}
 \item All GRBs within T$_{90}$ < 2.05 sec before the date of sGRB 170817A have been selected to initiate the analysis. This formed a sample of 635 sGRBs. 
\item The significant emission of sGRB 170817A was at the first peak $T_0-0.192$ sec to $T_0+0.064$ sec, where $T_0$ is the  starting time of the GRB. The spectral form for this period followed by a Comptonized model (Comp) function \citep{Gruber:2014iza} with three parameters, amplitude (\(A\)) in photon $\text{s}^{-1}$ $\text{cm}^{-2}\text{keV}^{-1}$, spectral index (\(\alpha\)), and peak energy  (\(E_{\text{peak}}\)) in keV, and $\epsilon_{\gamma,\text{piv}}$ is always kept at 100 keV for all the cases \citep{Poolakkil:2021jpc}. 
\begin{equation}
    F_{\text{Comp}}(\epsilon_{\gamma}) = A \left( \frac{\epsilon_{\gamma}}{\epsilon_{\gamma,\text{piv}}} \right)^{\alpha} \exp \left[ -\frac{(\alpha + 2)\epsilon_{\gamma}}{E_{\text{peak}}} \right]
\end{equation} 

\begin{equation}
    F_{\text{PL}}(\epsilon_{\gamma}) = A \left( \frac{\epsilon_{\gamma}}{\epsilon_{\gamma,\text{piv}}} \right)^{\alpha}
\end{equation}
sGRB 170817A has E$_{\text{peak}}$ and $\alpha$ value that represent a softer emission and consist of a smaller set in the catalog (\cite{Zhang:2017lpb}). Hence, out of the 635 sGRBs we selected the sGRBs below the upper limit of the 1$\sigma$ value of E$_{\text{peak}}$, which is $317$ keV. This now forms a set of 80 sGRBs. We have shown this process and cut-off values of E$_{\text{peak}}$ and $\alpha$ in Figure \ref{num_bin}.

\begin{figure*}
	\includegraphics[width=1.6\columnwidth, height=9.5cm]{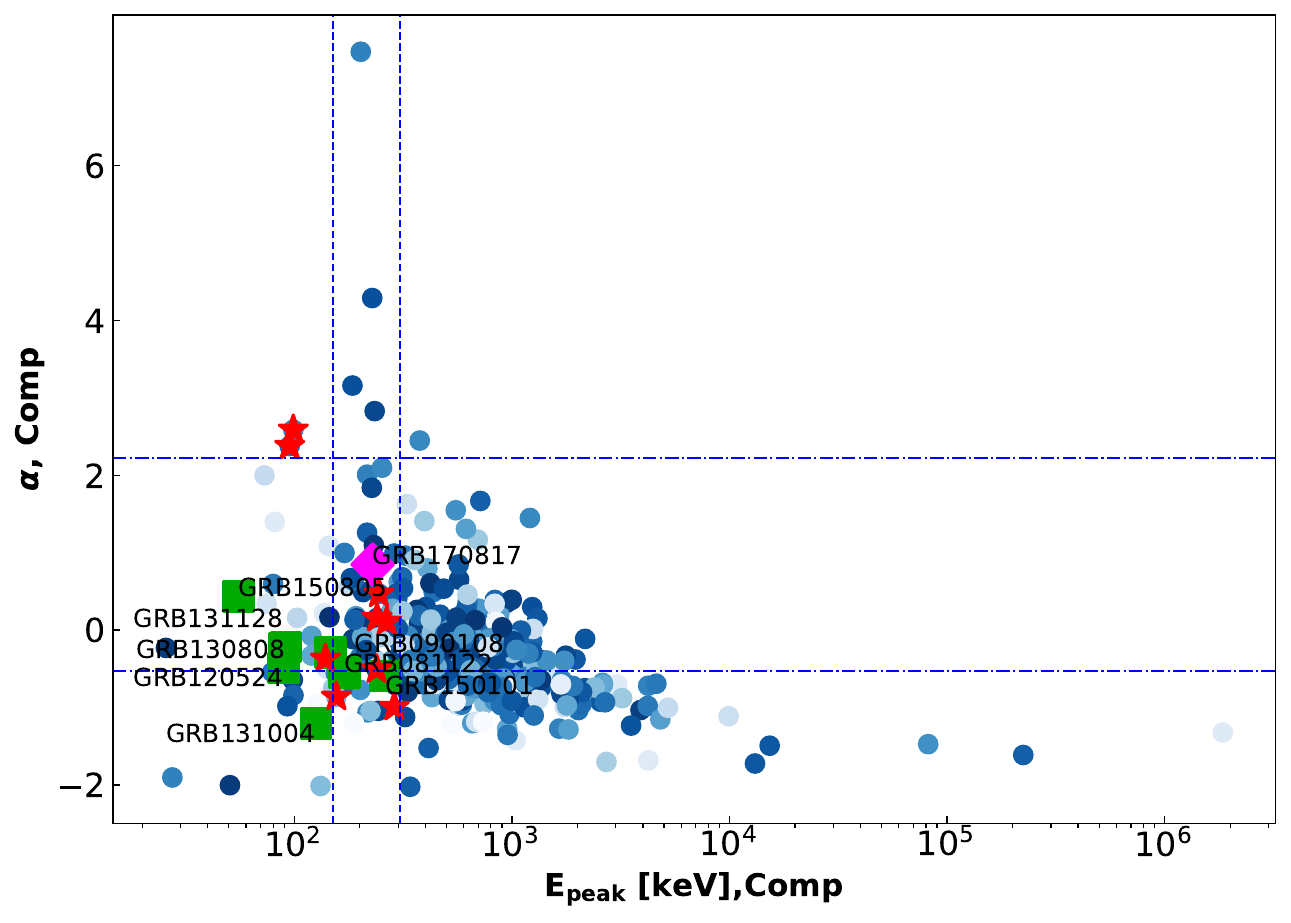}
    \caption{Scatter plot of Comp parameters from all archived sGRBs $T_{90}$ < 2.05 for the time-integrated spectrum. The sGRBs best fitted with PL and Comp are marked with light blue to dark blue $T_{90}$ values, where dark blue indicates higher $T_{90}$ values and lighter blue indicates lower $T_{90}$. The Comp parameters of sGRB 170817A are emphasized with a diamond ( magenta color), the $1\sigma$ error bars for $E_{peak}$ are shown in dashed lines, and for the $\alpha$ parameter, it is shown in dot-dashed lines. The eight possible sGRB 170817A-like bursts are marked with green squares. And sGRBs with known redshift are shown as stars (red color) (sGRB 150101B, sGRB 131004A are excluded).}
    \label{num_bin}
\end{figure*}
\item The further sampling is implemented following the fact that the second peak, $T_0+0.832$ sec to $T_0+1.984$ sec (\cite{Zhang:2017lpb}) of sGRB 170817A has a blackbody (BB) spectral form,

\begin{equation}
F_{\text{BB}}(\epsilon_{\gamma}) =  A\frac{ \epsilon_{\gamma}^2}{\exp^{\epsilon_{\gamma}/kT} - 1},
\label{bb}
\end{equation} \\

with the thermal component at energy kT $ =10.3 \pm 1.5$ keV for the second peak, where k is the Boltzmann constant. 
This sGRB has the $T_{90}$ start time and end time as \text{$T_{90}^{\text{start}}$}  = -0.192, \text{$T_{90}^{\text{end}}$}  = 1.856, respectively whereas the $T_{50}${\footnote{$T_{50}$ is the duration in which the central 50\% of the gamma-ray signal is detected.}} end time,  \text{$T_{50}^{\text{end}}$}  = 1.216, we chose to split the entire GRB time into two zones, Zone-A: \text{$T_{90}^{\text{start}}$} till  \text{$T_{50}^{\text{end}}$} and Zone-B: \text{$T_{50}^{\text{end}}$} till \text{$T_{90}^{\text{end}}$}, so that the two zones can accommodate the first peak and the second peak. We calculated the HRs, which are the ratios of the counts or flux in a high-energy band to that in a low-energy band, providing insight into the spectral characteristics of GRBs as a color index in the photometry. The HR provides insights into the energy distribution of the burst's emitted photons, which helpdistinguish between different types of GRBs and understandng their underlying physical mechanisms.
By normalizing the exposure-corrected count numbers in two energy bands, HR is computed. The energy bands are,
Low-energy band: 10-50 keV and
High-energy band: 50-300 keV, within these two energy bands we calculate the HR in the aforementioned two time zones for the above selected 80 sGRBs, following the definition of the HRs as, 
\begin{equation}
HR_{1} = \frac{\text{photons within } 50 \text{ to } 300 \text{ keV  }  \text{ in Zone-A } }
{\text{photons within } 10 \text{ to } 50 \text{ keV}   \text{ in Zone-A } }
\label{eqn-4}
\end{equation}
 and,
\begin{equation}
HR_{2} = \frac{\text{photons within } 50 \text{ to } 300 \text{ keV }  \text{ in Zone-B}}
{\text{photons within } 10 \text{ to } 50 \text{ keV  }  \text{ in Zone-B}}.
\label{eqn-5}
\end{equation}

The values of the two HRs for these 80 sGRBs are listed in table \ref{table-hr} and shown in a scattering plot in Figure \ref{mahal}. The detailed calculation of the HR from the Fermi-GBM data is explained in section \ref{tool}.

\item In a further step, we look for clusters within the HR scattering. However, before searching for such subgroups, we tried to eliminate the outliers to minimize bias. The outliers are identified using the  Mahalanobis distance ($D_m$) \citep{mahalanobis2018generalized}. It is a multivariate distance metric that measures the separation between a data point and the distribution's mean, considering correlations between the variables. In contrast to simple Euclidean distance, it incorporates the covariance matrix to account for the data's orientation and shape. $D_m$ \citep{de2000mahalanobis} is written as follows :
\begin{equation}
D_m(\mathbf{x}) = \sqrt{ (\mathbf{x} - \boldsymbol{\mu})^{T} \, \mathbf{S}^{-1} \, (\mathbf{x} - \boldsymbol{\mu}) }
\end{equation}
where x is a date point, $\mu$ is the mean of the distribution, and S is the covariance matrix of the dataset. Using this method,
we draw the $1\sigma$, $2\sigma$, and $3\sigma$ standard deviation ellipses with dotted, dot-dashed, and solid lines, respectively, as shown in figure \ref{mahal}. If the data point falls outside the $3\sigma$ range, it will be considered an outlier to the group. With the mentioned criteria, we find that sGRB 081101 and sGRB 120814 are the two outliers in our datasetas shown in Figurere \ref{mahal} as diamonds (red). 

\begin{figure}
\includegraphics[width=1.\linewidth]{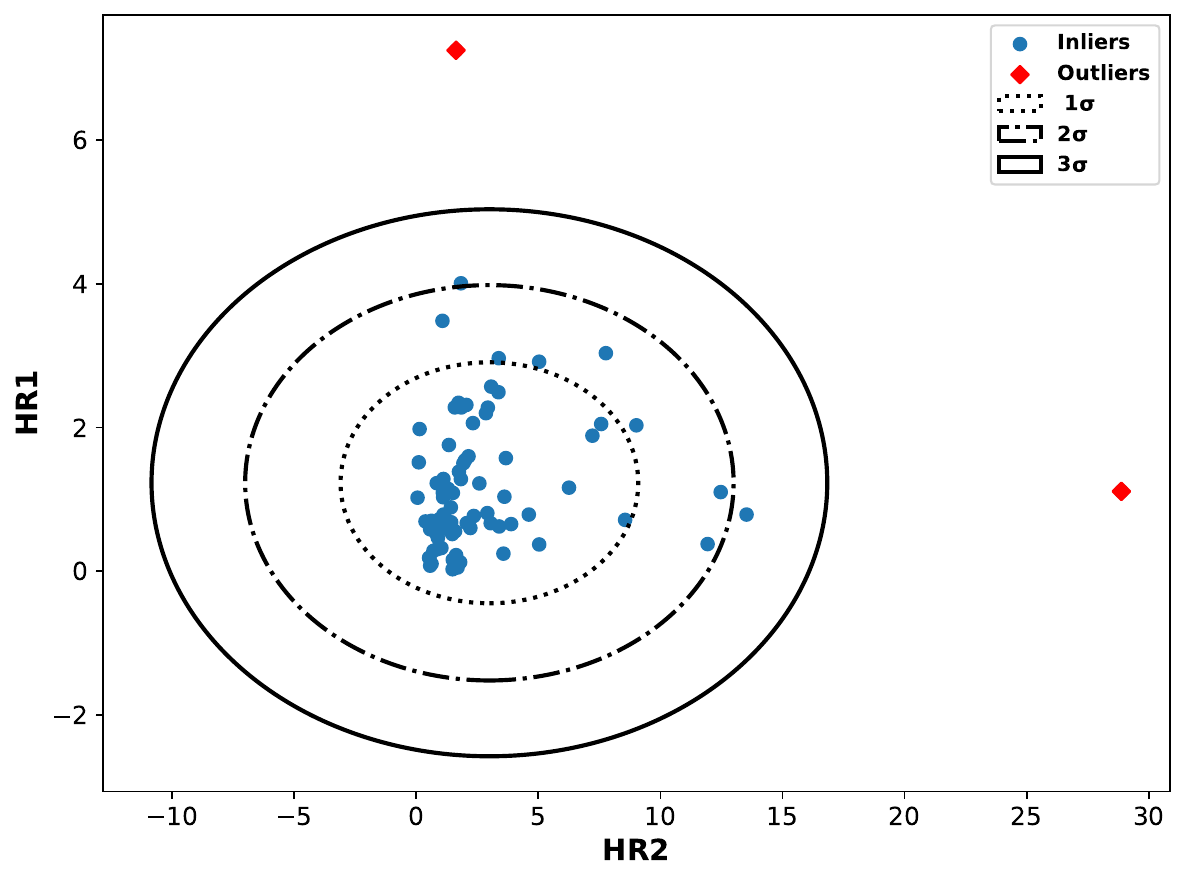}
    \caption{Distribution of data points in the $HR_1$ \& $HR_2$ parameter space is displayed via a Mahalanobis distance plot.  The ellipse show confidence contours that were obtained using the data's covariance. Dots (Blue color)represent the inliers (data used for further processing), while diamonds (red color) represente the outliers.}
    \label{mahal}
\end{figure}
\item In the scattering distribution of the 78 GRBs over their $HR_{1}$ and $HR_{2}$ values, we searched for subgroups in this step. We employed the K-means clustering algorithm for this purpose and found 6 possible clusters formed by the aforementioned 78 sGRBs. The details of the K-means adoption are explained in section \ref{ml}. Out of the 6 clusters, cluster 1 as shown in the figure \ref{fig-kmeans} contained sGRB 170817A along with the other 17 sGRBs, which are our fiOut of the 6 clusters, cluster 1, as shown in the figure \ref{fig-kmeans}, contained sGRB 170817A along with the other 17 sGRBs, which are our final list of sGRBs to do the full light curve and spectral analysis. above, we finally made alike sGRBs set based on the following two steps. a) presence of distinct double peaks in the light curve, a feature suggesting off-axis sGRBs, b) followed by the BB emission fitting for the Zone-B spectral with energy, kT $< 12$ keV, to look similar energy of sGRB 170817A. Out of these 17 sGRB, we found 8 sGRBs, sGRB 150805746, sGRB 150101641, sGRB 131128629, sGRB 131004904, sGRB 130808253, sGRB 120524134, sGRB 090108020, and sGRB 081122614 to satisfy the above-mentioned criteria and these details are mentioned in the table \ref{table-spec} and these 8 alike sGRBs are with a superscript of a solid star . We show the lightcurve of these 8 sGRBs along with the sGRB 170817A in figure \ref{lc-similar}, calling them as the alike sGRBs to sGRB 170817A. A detailed analysis method is presented in Section \ref{LC-SED}. We have shown the spectral fitting parameters for the Comp model in Zone-A and the BB model in Zone-B for all the 80 sGRBs collected after the second step of the data selection procedure, in table \ref{table-spec}. 

A schematic diagram of the above mentioned selection procedure of alike sGRBs to the sGRB 170817A, is demonstrated in figure \ref{flowchart}. 

\end{enumerate}
\begin{figure}
	\includegraphics[width=1.0\linewidth]{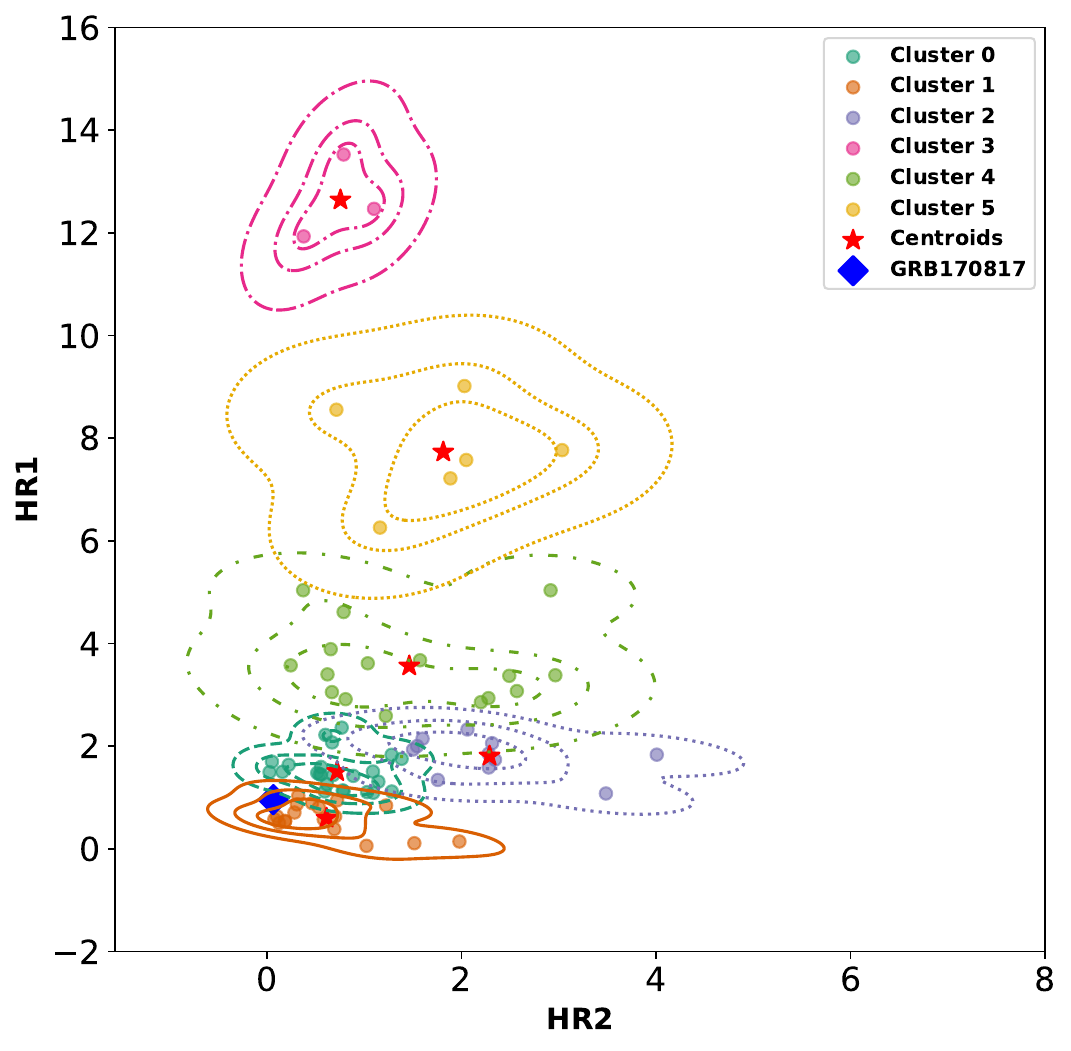}
    \caption{Contour based clustering ($HR_{1}$ v/s $HR_{2}$) at k = 6, each cluster represented by a distinct color. Contour lines indicate the density distribution of points within each cluster, which visually represents the data's structure, highlighting regions with higher concentrations of data points.  Stars (red color) are centroids, the diamond ( blue color) is sGRB 170817A.}
    \label{fig-kmeans}
\end{figure}

\begin{figure}
\begin{tikzpicture}[node distance=1.5cm]

\tikzstyle{startstop} = [
    rectangle,
    rounded corners,
    minimum width=3cm,
    minimum height=1cm,
    text centered,
    draw=black
]

\tikzstyle{process} = [
    rectangle,
    minimum width=3cm,
    minimum height=1cm,
    text centered,
    draw=black
]

\tikzstyle{arrow} = [thick,->,>=stealth]

\node (in1) [startstop, text width=8cm, align=center]
{\textbf{Data Collection:} Fermi-GBM catalog sGRBs with T$_{90}$ $\textless$ 2.05 sec till 17\textsuperscript{th}
 August 2017, \textbf{635} sGRBs are collected};

\node (pro1) [process, below of=in1, text width=8cm, align=center]
{$\bm{F_{1}:}$ sGRBs within  $1\sigma$ values of $E_{peak}$ and $\alpha$  for Comp model of sGRB 170817A, leaving \textbf{80} sGRBs };

\node (pro2) [process, below of=pro1, text width=8cm, align=center]
{$\bm{F_{2}:}$ Hardness ratio's $HR_1$, and $HR_2$ are calculated for above \textbf{80} sGRBs in two time Zones};

\node (pro3) [process, below of=pro2, text width=8cm, align=center]
{$\bm{F_{3}:}$ \textbf{Two} outliers in the $HR_1$, $HR_2$ parameter space identified using Mahalanobis distance method};

\node (pro4) [process, below of=pro3, text width=8cm, align=center]
{$\bm{F_{4}:}$ \textbf{6} Subgroups resulted using K-means clustering, \textbf{17} sGRBs found in the subgroup of sGRB 170817A};

\node (pro5) [process, below of=pro4, text width=8cm, align=center]
{$\bm{F_{5}:}$ \textbf{17} sGRBs fitted with BB spectral model in Zone - B, \textbf{8} sGRBs idenfied with kT $< 12$ keV as alike to sGRBs 170817A };

\draw [arrow] (in1) -- (pro1);
\draw [arrow] (pro1) -- (pro2);
\draw [arrow] (pro2) -- (pro3);
\draw [arrow] (pro3) -- (pro4);
\draw [arrow] (pro4) -- (pro5);

\end{tikzpicture}
\caption{Workflow illustrating our process of finding out the twin sGRBs alike sGRB 170817A. \textbf{F} refers to the various filtering phases followed.}
\label{flowchart}
\end{figure}
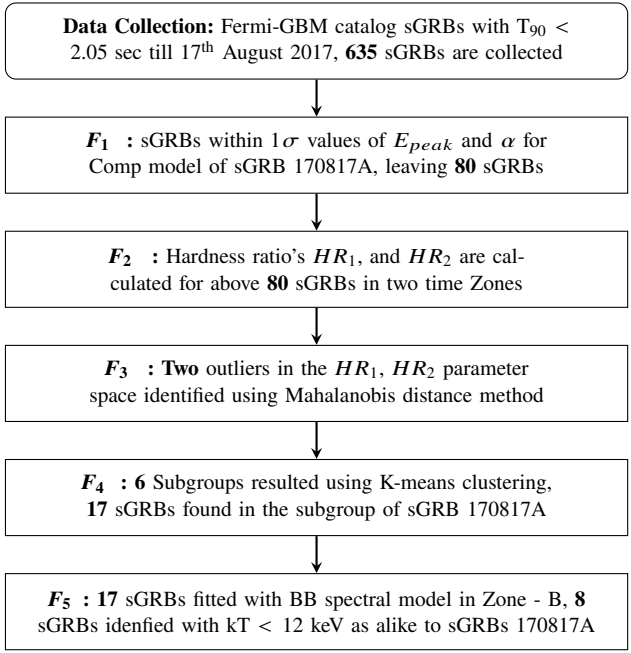

\section{HR\textsubscript{1} and HR\textsubscript{2} Calculations}
\label{tool}
As defined in section \ref{archive}, the HR\textsubscript{1} \& HR\textsubscript{2} are the ratio of photon counts observed within energies 50 keV to 300 keV and 10 keV to 50 keV in time Zone-A and Zone-B, respectively. These parameters generally indicate the spectral transition of the GRBs through the total time of emission. We calculated the HR\textsubscript{1} \& HR\textsubscript{2} for the 80 sGRBs selected in the second step
of the data selection criteria, using a forward-folding spectral analysis software RMFIT -- v4.3.2 package\footnote{https://fermi.gsfc.nasa.gov/ssc/data/p7rep/analysis/rmfit/}} (\cite{mallozzi2005rmfit}).

The main purpose of RMFIT is to obtain the best-fit parameters for a chosen model from data files that include observed count rates (PHA files: photon count rate files), a matching detector response matrix file (RMF), and an auxiliary response file (ARF). For our case, we used Time-Tagged Event (TTE) data files, which have 128 energy channels with varying time bins. We utilized the graphical user interface of RMFIT to time- or energy-bin the files and determine the background, then saved both the signal and background files in ASCII format. From these files, we calculated the HR$_{1}$ and HR$_{2}$ using the two most intense NaI detectors (Thallium-activated sodium iodide detector) mentioned in the burst catalog. Using the ASCII files generated from the TTE files, all the photon counts over the selected energy range will be added for the chosen two NAI detectors after subtracting the background. Then we calculated the $HR_1$ and $HR_2$ for the 80 sGRBs as selected in the second step of the data selection procedure, and listed the values in table \ref{table-hr} with the obtained photon counts along with the time zones, selected detectors, and redshift values  (if available) of all 80 sGRBs. 

\subsection{Subgroups in HR\textsubscript{1} vs HR\textsubscript{2} using K-Means Clustering} \label{ml}
An in-depth understanding of the uniqueness or similarity with sGRB 170817A can be obtained by looking at the HR\textsubscript{1} versus HR\textsubscript{2} distribution of the 78 selected sGRBs. One of the most widely used unsupervised clustering techniques is K-means clustering (\cite{lloyd1982least,macqueen1967some, na2010research,jin2011k}). Here, we used this technique to search for further sub-grouping in the final dataset. The selection of the optimal number of clusters (K) is one of the primary tasks. Although determining K is not necessarily a clear-cut exercise, especially when the data is unlabeled. However, heuristic approaches coupled with a good understanding of the data set and its features are sufficient to determine a range of optimal values of K. For this task, numerous initializing strategies are available. We used the elbow method (\cite{syakur2018integration,umargono2020k,cui2020introduction}) to find out the optimal number of K. For that, we have calculated the inertia  (sum of the squared distances of each point to its assigned centroid). Inertia of a cluster can be calculated as the,

\begin{equation}
    Inertia = \sum_{i=1}^{N} \sum_{j=1}^{K} \min ( \| x_i - \mu_j \|^2 )
    \label{eqn-8}
\end{equation}

where $x_i$ are the data points in the cluster, $\mu_i$ is the centroid of the cluster, and N \& K are the set of data points and the number of clusters, respectively. In our case, we have calculated the inertia value for various K values from 2 to 35. The value of the minimum inertia calculated using equation \ref{eqn-8} is our optimal K, which is called the elbow (knee point) \citep{del2016study,antunes2018knee}.

\begin{figure}
	\includegraphics[width=1.\linewidth]{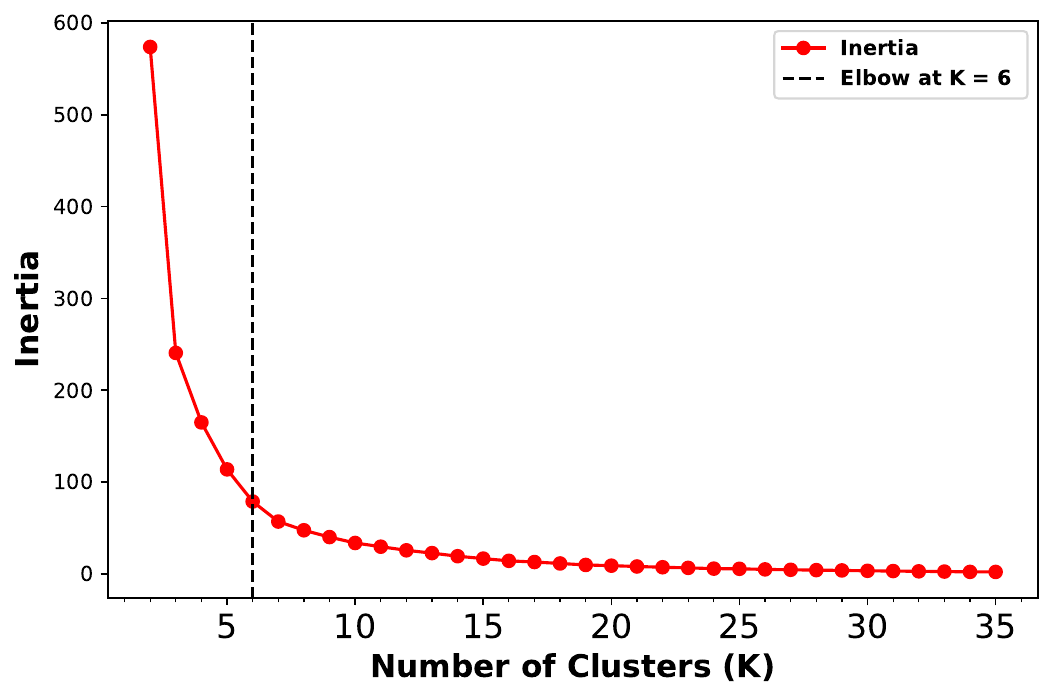}
    \caption{Variation of the inertia with the number of clusters. The black dashed line is the kneedle point representing the maximum curvature, corresponding to the  optimal K number}
    \label{elbow}
\end{figure}

Although it is ideal to discover the point at which the line bends (forming an elbow shape) to obtain an accurate estimation of the K value. Most real data sets have a smooth curve, making it impossible to find the optimal K value. This is true for our case, as shown in figure \ref{elbow}. Since the elbow feature is not clear for our data set, therefore, to find the optimal K value, we used the Kneedle algorithm \footnote{\url{https://www.kaggle.com/code/kevinarvai/knee-elbow-point-detection}} \citep{satopaa2011finding,qumsiyeh2023utilizing,schubert2023stop,fok2024deep}). This algorithm is a generic tool that uses K versus the inertia curve to detect the optimal K value for clustering tasks. It relies on the mathematical concept of curvature to measure how much a curve deviates from a straight line. Here, we identify the point of maximum curvature (finding the local maxima in a set of points), which occurs when the curvature decreases and the curve becomes flatter. Using this method, we found that the ideal number of clusters, K, is 6, shown with a vertical dashed line in figure \ref{elbow}. Furthermore, the kneedle point does not vary up to K = 35, since 35 is a huge number for our dataset, so we ignored it. 

Once the optimal number of clusters is known, we proceed with calculating the distance from each data object to all the centers for the K-number of clusters.

In general, euclidean distance between two data points is calculated as follows: suppose we have two data points a and b in euclidean space of n-dimensional length, where  a and b are \( a = (a_1, a_2, \dots, a_n), \quad b = (b_1, b_2, \dots, b_n) \)
 then ecludian distance d(a,b) can be written according to \citep{na2010research}.
\begin{equation}
    {\rm {d}}({{\rm {a}}_{\rm {i}}},{{\rm {b}}_{\rm {i}}}) = {\left[{\sum\limits_{i = 1}^n {{{\left({{a_i} - {b_i}} \right)}^2}} } \right]^{1/2}}
    \label{eqn-7}
\end{equation}

In K-means clustering, calculating the criterion function (J) or the cost function \citep{li2012clustering,na2010research} is a crucial step in labeling the data points with a particular cluster. It is defined as the sum of squared distances between data points with their respective centroids in all the K-number of clusters; 
\begin{equation}
  {\rm {J}} = \sum\limits_{i = 1}^K {\sum\limits_{x \in {C_i}} {{{\left| {x - {x_i}} \right|}^2}} } 
  \label{eqn-6}
\end{equation}\\
 This parameter is calculated for several iterations.
In each iteration, the mean distance of the cluster points from their respective centroids is calculated, followed by the relocation of the centroids. This method continues until the cluster centroids remain unchanged for two consecutive successful iterations or the value of the criterion function (or the cost function) falls below a specific threshold value. This point is considered to be the case where the algorithm has converged or stabilized. In this work, this threshold value is set to $10^{-4}$.
 We clustered our data using K = 6 to find clear patterns in the final dataset and examined the resulting groups. Figure \ref{fig-kmeans} visually represents the results of the clustering, with contours that emphasize the density distribution of the data points and different colors for each cluster. Our root sGRB 170817A, visualized with the blue diamond, lies in cluster 1, which is brown in color; there are a total of 17 sGRBs in cluster 1. Interestingly, 8 out of 17 sGRBs in the cluster satisfies the selection criteria mentioned above.
 This clustering result strongly suggests us that these sGRBs share comparable traits, making the cluster especially pertinent for further research.

\subsection{LIGHT-CURVES \& SPECTRAL ANALYSIS} \label{LC-SED}
This time, to precisely understand the similarity within the sGRB subgroup above, we perform temporal and spectral analysis using the GBM data. 
We discussed previously that the light curve of sGRB 170817A has two distinct features in both the temporal evolution and spectral form. 
For completeness, we analyzed these 80 sGRBs along with sGRB170817A using RMFIT 
analysis with the publicly available Fermi-GBM TTE data. 

In the data collection step for the analysis, we selected only the two NaI detectors for which the position of the GRB is within $60^{\circ}$ (Good geometry) and has the highest intensity, as discussed in section \ref{tool}. Data are collected for energy channels 8 keV to 1 MeV, excluding channels around $\sim 30 - 40$ keV \citep{Kaneko:2006ru,2025NatCo..16.2668J} because of iodine K-edge at 33.17 keV for NaI detectors. In order to subtract the background from the GBM data, we fitted data from two time interval values before and after the GRB prompt emission (around the selected time interval, i.e., T$_{90}$) with a polynomial as the background model (for lightcurve analysis, see :
\citep{Salafia:2016wru,DelVecchio:2016vjn,2024arXiv241113242K}. We defined the signal above the background by interpolating this polynomial around the GRB prompt phase.

\begin{figure*}
\centering
\begin{tabular}{@{}ccc@{}}
  \includegraphics[width=0.33\textwidth]{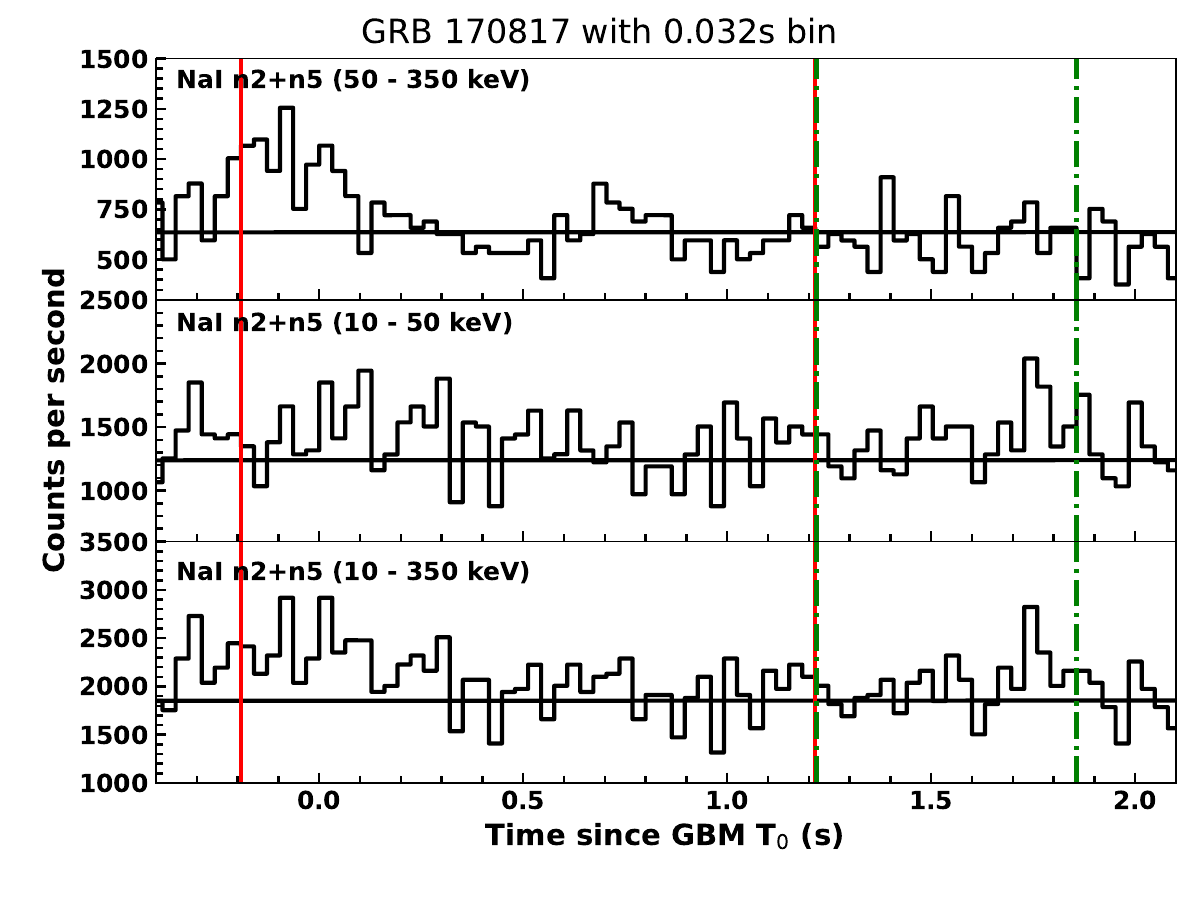} &
  \includegraphics[width=.33\textwidth]{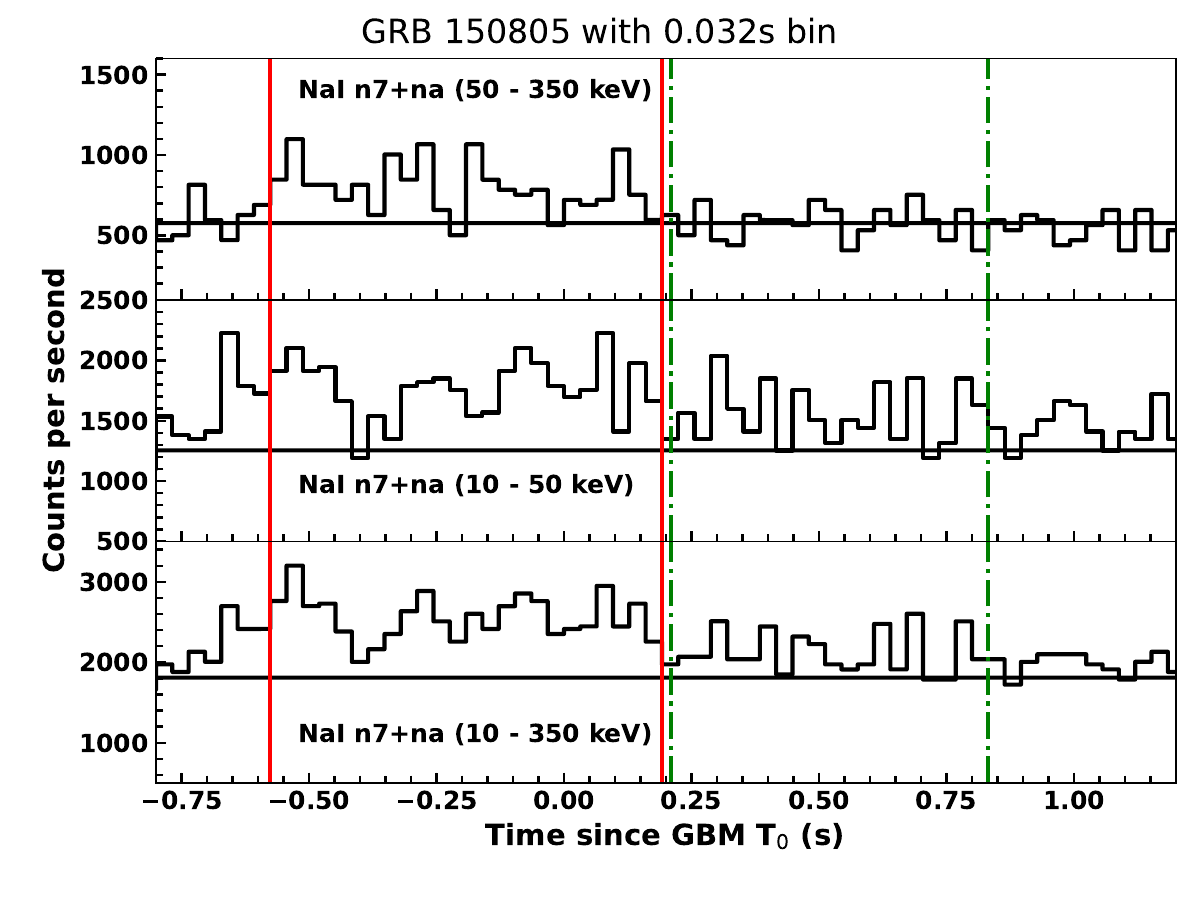} &
  \includegraphics[width=0.33\textwidth]{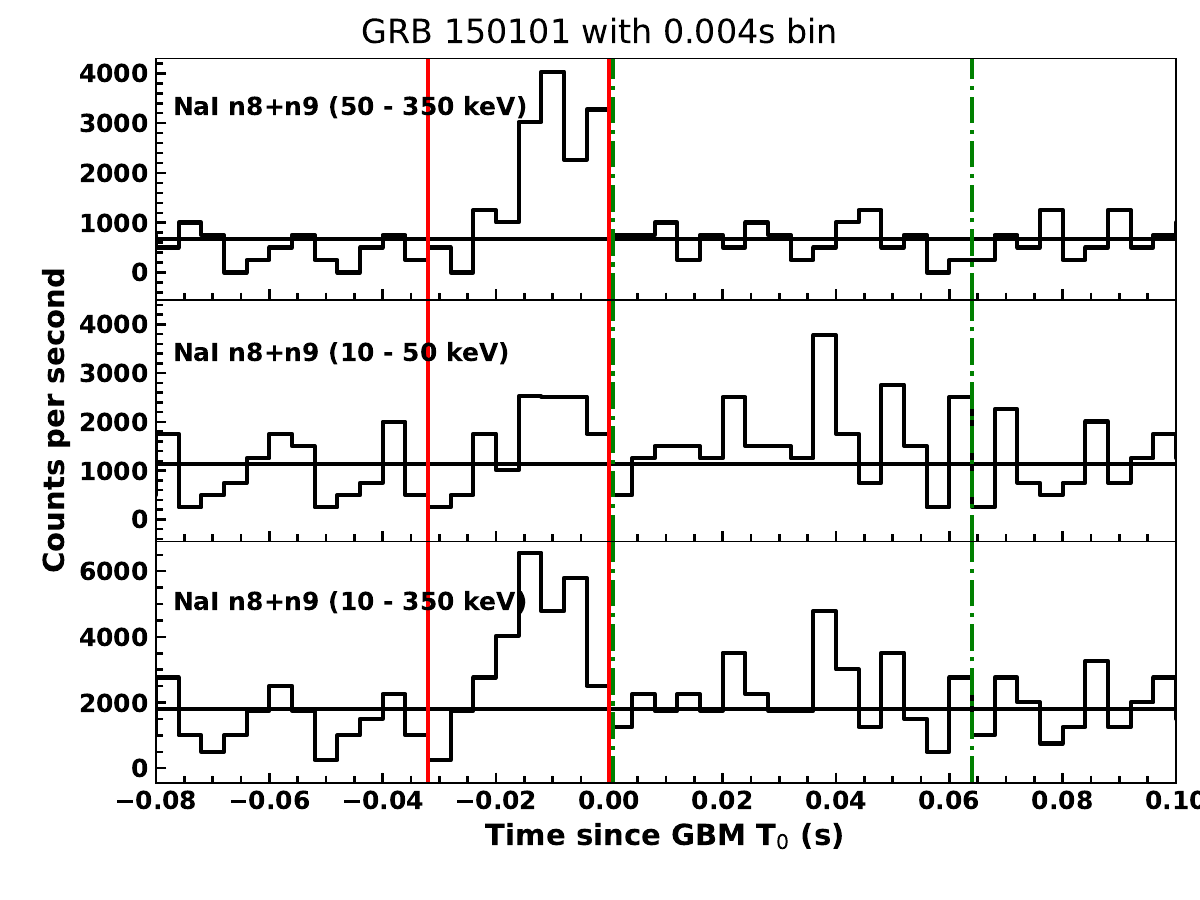} \\ [1ex]
  
  \includegraphics[width=0.33\textwidth]{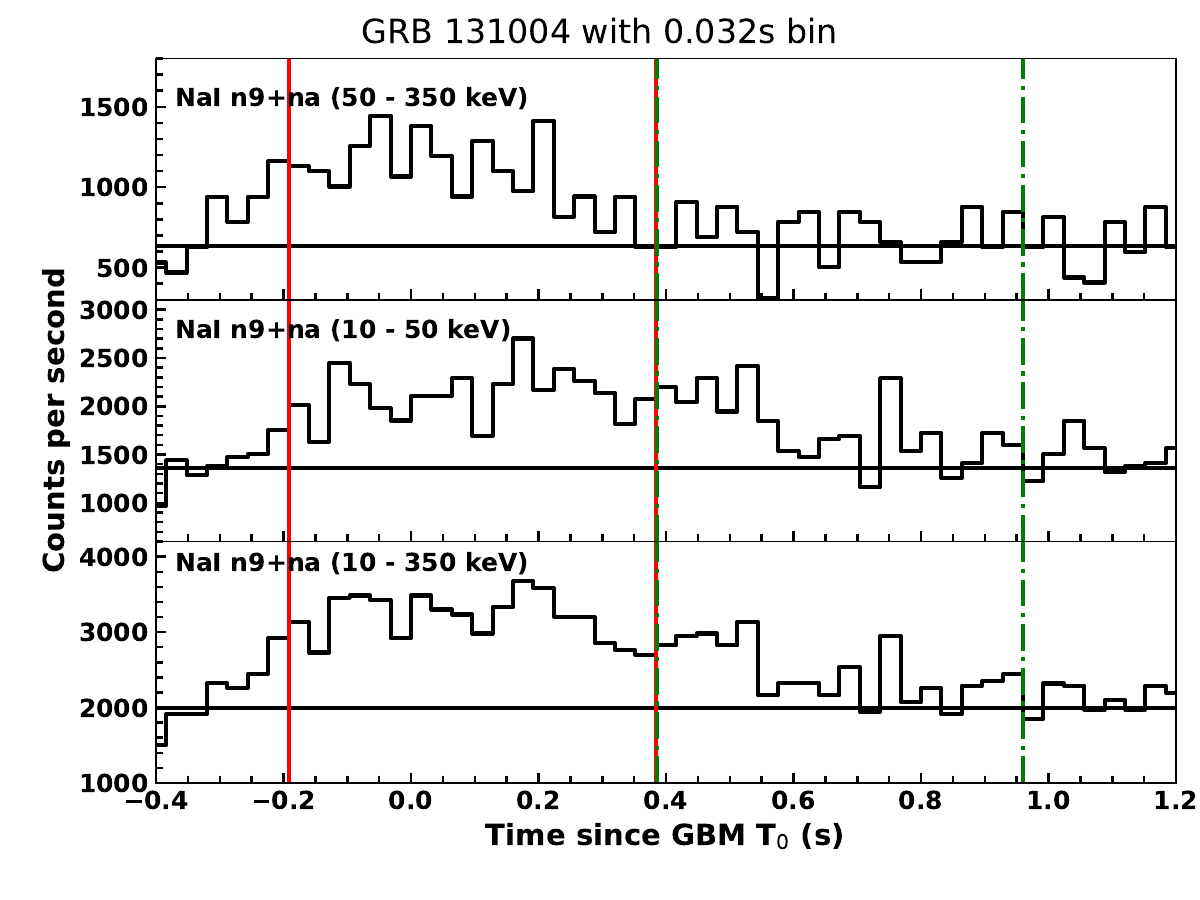} &
  \includegraphics[width=0.33\textwidth]{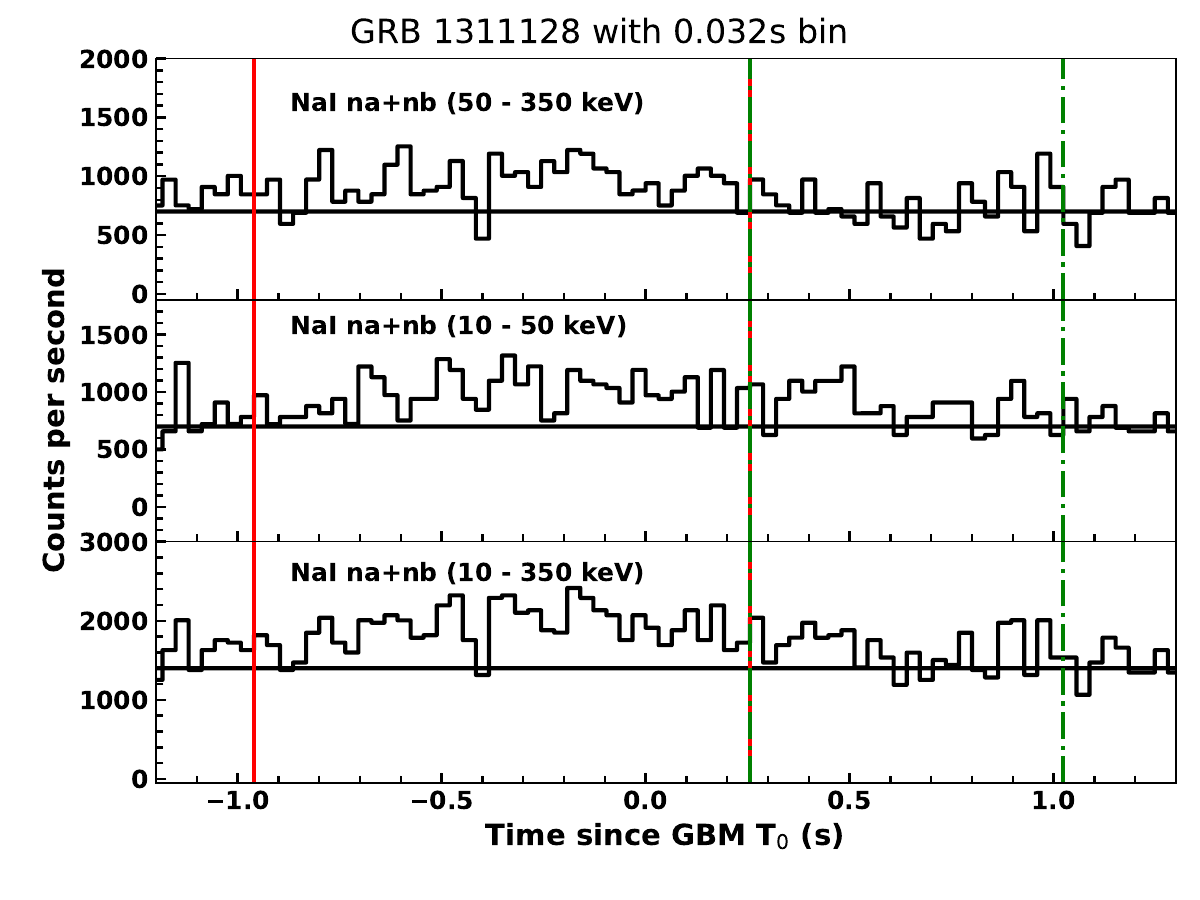}&
  \includegraphics[width=0.33\textwidth]{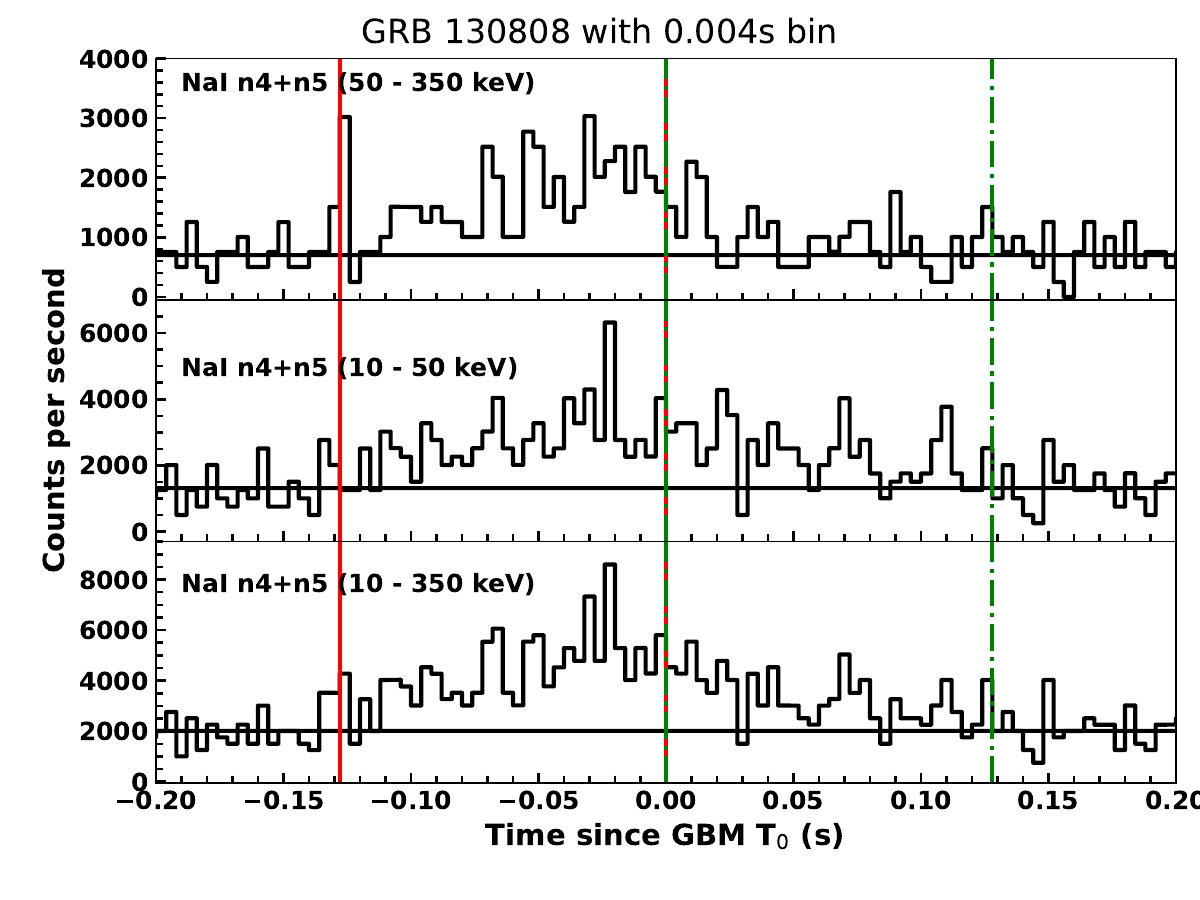} \\ [1ex]
  \includegraphics[width=0.33\textwidth]{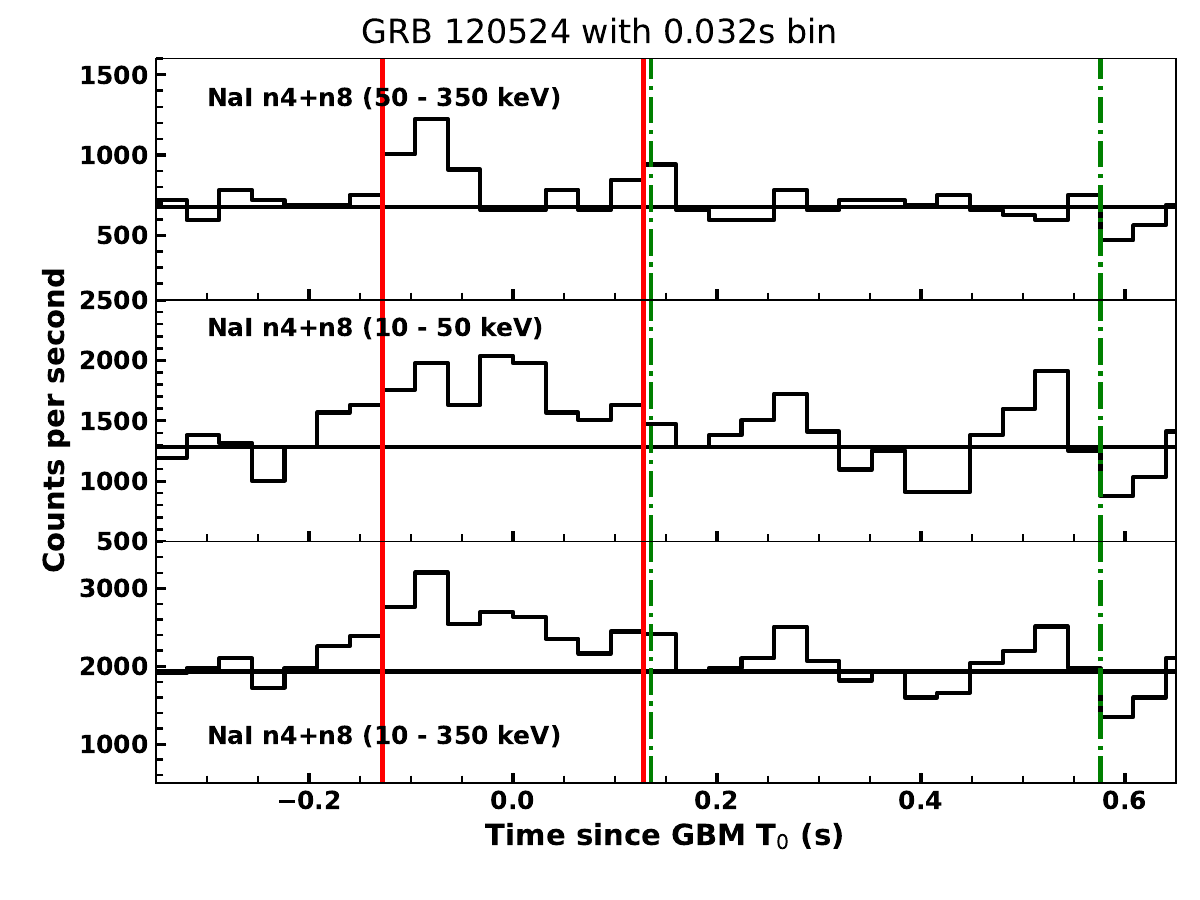} &
  \includegraphics[width=0.33\textwidth]{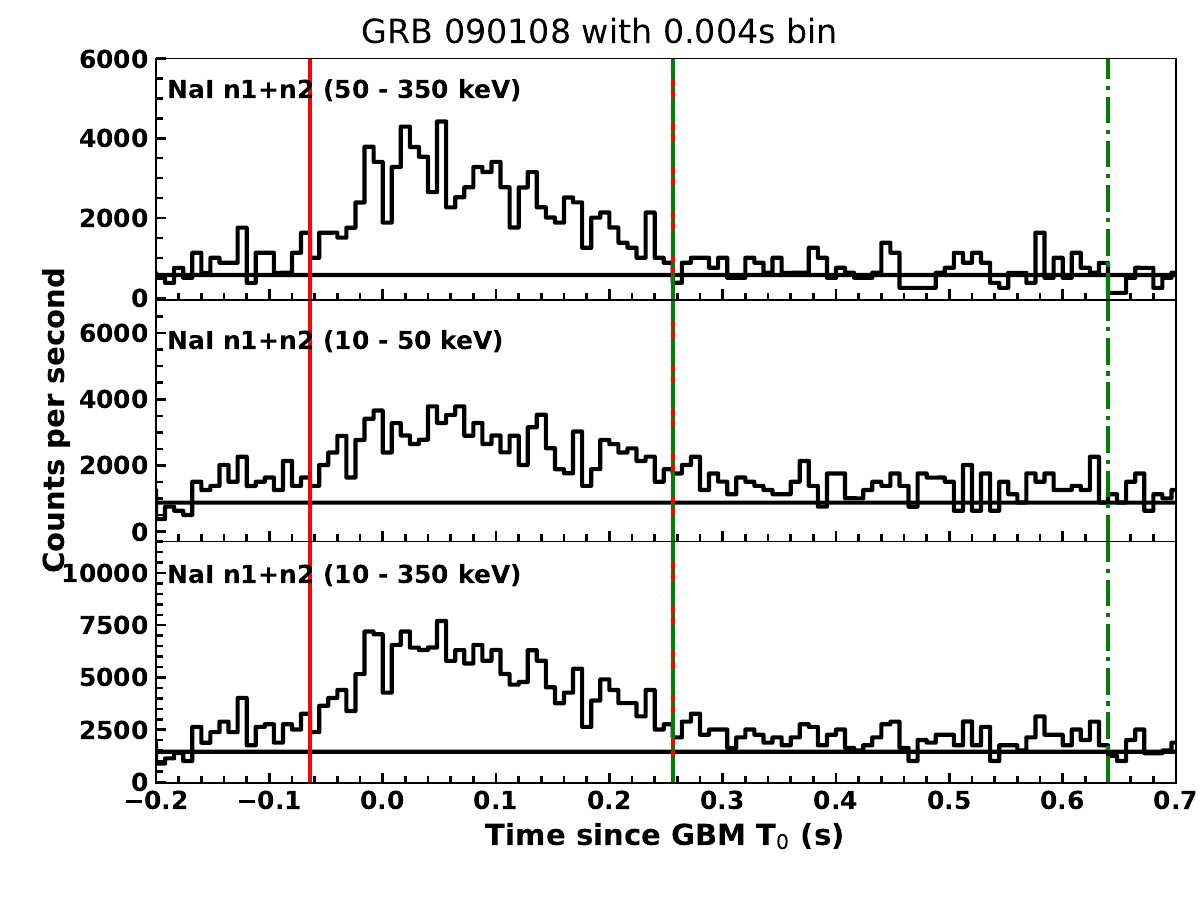} & 
  \includegraphics[width=0.33\textwidth]{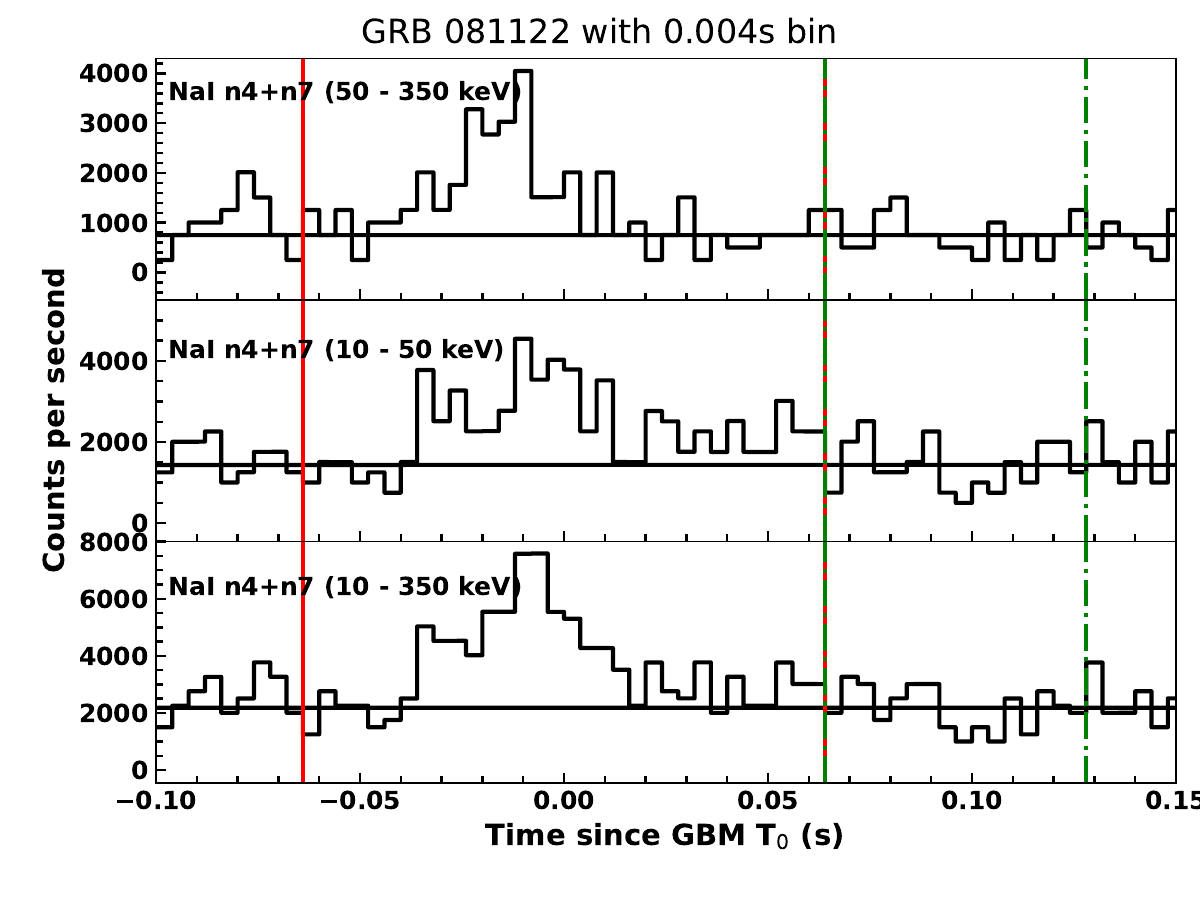} \\ [1ex]
\end{tabular}
\caption{sGRBs similar to sGRB 170817A. The order of the light curves is the following: sGRB 170817A, sGRB 150805, sGRB 150101, sGRB 131128, sGRB 131004, sGRB 130808, sGRB 120524, sGRB 090108, and sGRB 081122. All the light curves are produced in the three energy ranges 50-350 keV, 10-50 keV and 10-350 keV with the binning size of 0.032 sec and 0.004 sec. The first and second peak time zones are shown with a solid (red) line and a dot-dashed (green) line, respectively. The light curve is the addition of the counts from the NAI detectors used for analysis. The horizontal solid (black) line is the background. }
\label{lc-similar}
\end{figure*}

We studied the light curve of the 8 sGRBs selected following section \ref{archive}, similar to sGRB 170817A. For reference, we have also shown the same analysis result for sGRB 170817 figure \ref{lc-similar}. The figure includes LCs for different energy ranges 10-350 keV, 10-50 keV, 50-350 keV, from bottom to top. We show the two time zones with red (solid lines) and green (dot-dashed lines), respectively. 

We performed the integrated spectral analysis for the two zones. To have an unbiased comparison, we have also analyzed sGRB 170817A using the same method, and the corresponding results are listed in table \ref{table-spec}, which contains spectral analysis parameters for all the 79 sGRBs initially collected. We have highlighted the 8 sGRBs mentioned above with bold letters and superscript-star symbol in the table \ref{table-spec}.

According to our analysis, the spectrum of the Zone-A of sGRB 170817A follows the Comp model with $E_{\text{peak}}= 176.3 \pm 85.7$ keV and $\alpha = 0.93 \pm 0.16$. The Comp spectral Castor C statistic (C-stat) improvement \citep{1979ApJ...228..939C} of 3.21 therefore, has a chance of occurrence $\sim 7.31 \times 10^{-2}$ compared to PL. The second peak spectrum follow BB with kT$=9.23 \pm 1.83$. The C-stat spectral value of BB improves by 8.91 compared to the PL fit corresponding to a chance of occurrence $\sim 2.83 \times 10^{-3}$. The number of variable parameters in BB is the same as for the PL function, and hence the result gives a statistically significant choice for the spectra to be BB radiation. We have also checked our analysis considering the same time zones for sGRB 170817A as in \citep{von2019fermi} and found negligible variation in results. 

In addition, references \citep{burns2018fermi,Troja:2017nqp} also claimed sGRB 150101B as a luminous version of sGRB 170817A with a redshift of 0.1341. This GRB also belongs to one of the 8 sGRBs we have listed. Spectral analysis showed that the first zone from $T_0 -0.016$ to $T_0 -0.0$ s follows Comp with E$_{\text{peak}} = 162.8$ and $\alpha$ = 0.96 over PL with a chance of occurrence only 0.05. And the soft tail within $T_0 -0.0$ to $T_0 \pm 0.064$ has a spectral fit of BB with kT$ = 10.58$ preferred over PL with a chance of occurrence 0.03.
\section{RESULTS and Discussions}\label{result}
We have studied the sGRBs observed by Fermi-GBM to identify GRBs similar to the only significant EM counterpart of the GW event observed to date, sGRB 170817A. It is one of the lowest-luminescence GRBs observed to date, with a distance of only $\sim 42$ Mpc. This sGRB shows unique and significant features in both the light curve and the spectral form of the observation. The primary emission (first zone) follows the Compton emission form, whereas in the later stage (second zone), the emission is dominated by BB radiation. Importantly, the two zones are separated by a distinctive time difference of 0.576 sec (as shown in table 3 of \citep{2017ApJ...848L..14G}. The time resolution for such a gap may not be identified for other GRBs. Hence, in our study, we focused solely on the visibility of the two distinct spectral forms. Our search resulted in eight sGRBs that may have alike origin, in the Fermi-GBM catalog observed from 17 July 2008 to 17 August 2017. 

The statistical steps of this study are: We searched for subgroups in the distribution of HR parameter space of the two time zones, Zone-A (\text{$T_{90}^{\text{start}}$} to  \text{$T_{50}^{\text{end}}$} ) and Zone-B (\text{$T_{50}^{\text{end}}$} to \text{$T_{90}^{\text{end}}$}) for all the GRBs similar in the spectral form to sGRB 170817A in their Zone-A. We have used K-means statistics for subgrouping. Although we employed the kneedle method as explained in the section \ref{ml} for identifying the number of subgroups in the data, we utilized the elbow method. The last part of our search includes the spectral analysis and light curve analysis of sGRBs, and we further identified the softer BB-emitting sGRBs in Zone-B, using RMFIT analysis. The light curves of the resulting eight events are shown in the figure \ref{lc-similar}.
 
Out of the eight events, we found that two have a narrow second peak time period $\delta t_{2} = \text{$T_{90}^{\text{end}}$} - \text{$T_{50}^{\text{end}}$ }$ compared to the other six. Like for sGRB 150101B, and sGRB 081122, the second peak time period $\delta t_{2} = 0.064$ sec. Other sGRBs, such as sGRB 150805, sGRB 131128, sGRB 130808, sGRB 131004A, sGRB 120524, and sGRB 090108 have $\delta t_{2}$ values of 0.64, 0.768, 0.128, 0.576, 0.448, and 0.384 sec, respectively adding the sGRB 170817A has the $\delta t_{2}= 0.64$ sec. 

For sGRB 150101B, and sGRB 131004A, with known redshift, we observed E$_{\text peak} = 162.8$ keV and $107.3$ keV, respectively from our analysis in the Zone-A. The remaining sGRBs, sGRB 150805, sGRB 131128, sGRB 130808, sGRB 120524, sGRB 090108, and sGRB 081122 have measured $E_{peak}$ at 76.35 keV, 79.41 keV, 83.05 keV, 63.65 keV, 140.3 keV, and 143.8 keV, respectively, in zone A. The E$_{\text{peak}}$ of sGRB 150101B, sGRB 131004A, sGRB 090108, and sGRB 081122 are similar to our root sGRB 170817A, where other GRBs are having a softer spectrum than the GRB 170817A. \\

\begin{figure}
	\includegraphics[width=1.0\linewidth]{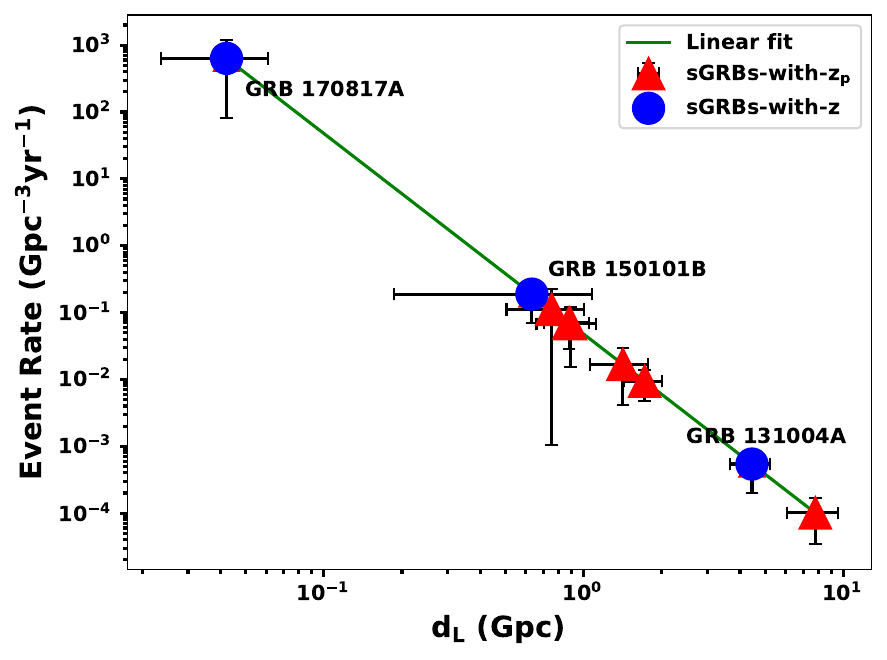}
    \caption{ Luminosity distance vs Event rate of 8 sGRBs we found through our analysis with sGRB 170817A; circles (blue color) represent the sGRBs with measured redshift (z) and triangles (red color) indicate the sGRBs with calculated pseudo-redshift ($z_{p}$), green line is linear interpolation function fitted with data.}
    \label{event-rate}
\end{figure}

In the case of GRB 170817A the integrated energy fluence for the Comp spectrum during Zone-A is $1.20 \pm 0.4 \times 10^{-7}$ erg cm$^{-2}$ and the BB spectrum during Zone-B has value $2.66 \pm 0.23 \times 10^{-8}$ erg cm$^{-2}$. The BB-spectrum for the second peak has energy (kT) $9.23  \pm 1.83$ keV. While for the other sGRBs, sGRB 150805, sGRB 150101B,  sGRB 131128, sGRB 130808, sGRB 131004A, sGRB 120524, sGRB 090108, and sGRB 081122 have energy (kT) values of $9.21$ keV, $9.82$ keV, $10.41$ keV, $12.00$ keV, $7.76$ keV, $7.60$ keV, $10.73$ keV, $11.26$ keV, respectively. The spectral analysis results of the sGRBs for both Zone-A and Zone-B are listed in the table \ref{event-rate}. A similar dedicated search has also performed in \citep{von2019fermi}. Their result yields a total of 13 candidates, including GRB170817A and the earlier reported similar burst, GRB 150101B \citep{Burns:2018qwx}. Their search is focused on peak searches in the light curve, emphasizing the presence of a valley between two peaks. We found no correlation in our resulted sample with the nearby sample of sGRBs resulting from BNS merger events from reference \citep{von2019fermi}, except GRB 150101B, since such a valley does not exist in the similar candidate GRB 150101B. Additionally, all the claimed candidates in reference \citep{von2019fermi} could not pass our selection criteria F1 (see figure \ref{num_bin}). Our search focuses on looking for spectral forms defined by the observed $T_{90}$ and $T_{50} $ of the sGRBs. 

Observation of sGRB170817A surprised the high-energy astrophysics community with its rate, $R= 680 - 1300$ $Gpc^{-3}$ $yr^{-1}$. We attempted calculating the event rate considering the sample of the 8 sGRBs selected as twining sGRB 170817A. Unfortunately, only two events out of the sample have measured redshift, sGRB 150101B and sGRB 131004A. The non-availability of spectroscopic measurements for the other 6 sGRBs leads to uncertainty in the distance of these events. Hence we followed the concept of pseudo-redshift ($Z_{p}$) as developed in reference \citep{2003A&A...407L...1A,2006AdSpR..38.1338P}, known as the Amati relation \citep{2002A&A...390...81A}, but widely used for long GRBs (LGRBs). In this relation, the distribution of E$_{\text{peak}}$ (here, as calculated in Zone-A) and E$_{\text{iso}}$ in the source rest frame (E$_{\text{peak}}^i = (1+z) E_{\text{peak}}$) has linear relation. 
\begin{equation}
E_{\text{peak}}^{\text{i}} = k \left( \frac{E_{\text{iso}}}{10^{52}\,\mathrm{erg}} \right)^{m} \, \mathrm{keV},
\label{amati}
\end{equation}
\begin{figure}
	\includegraphics[width=0.9\linewidth]{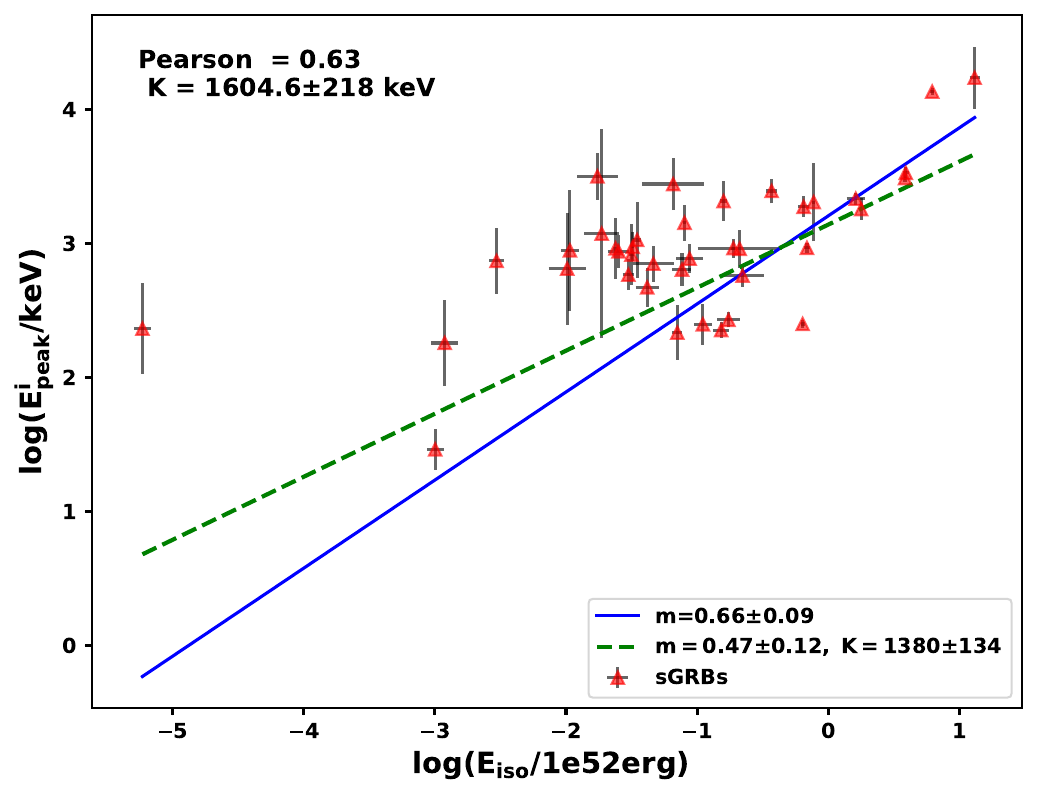}
    \caption{Correlation of $E_{iso}$ with $E_{peak}$ of Fermi-GBM sGRBs with known redshift till 17 August 2017.}
    \label{m-k}
\end{figure}
here, k and m are the fitting constants, and $E_{\text{iso}}$ is isotropic energy which can be written as, $E_{\text{iso}} = \frac{4 \pi d_{L}^{2}}{1+z} f $ erg,  $d_{L}$ is luminosity distance, $f$ is the energy fluence with unit $\text{erg cm}^{-2}$. Using cosmological constants \citep{WMAP:2008lyn}  $\Omega_M$ = 0.27,\ $\Omega_\Lambda$ = 0.73,\ $H_{0}$ = 70\ $\mathrm{km\,s^{-1}\,Mpc^{-1}}$ one can calculate $ d_{L} = \frac{(1+z)c}{H_{0}}\left[ \int_{0}^{z} \frac{dz'}{\sqrt{\Omega_{M}(1+z')^{3} + \Omega_{\Lambda}}}\right]$.

The Amati relation has been well established for LGRBs. However, such linearity could not be verified for sGRBs \citep{2022Ap&SS.367...74Z}. We have also shown the distribution plot of E$_{\text{peak}}$ vs E$_{\text{iso}}$ for all the sGRBs observed by Fermi-GBM with redshift up to the detection of sGRB 170817A, in figure \ref{m-k}. A Pearson's correlation coefficient (r) of 0.63  could be established on fitting a linear line similar to Amati relation, to these 40 sGRBs.  the linear relation parameters are m = $0.66 \pm 0.09$ and k = $1604.6 \pm 218$. This figure also clarifies, adding more data compared to reference \citep{2022Ap&SS.367...74Z}, but still does not improve the result.   

Recently,  \citep{2022Ap&SS.367...74Z} reported statistically significant linearity in the Amati relation (E$_{\text{peak}}$ vs E$_{\text{iso}}$) using a joint sample of Fermi-GBM and Swift sGRBs. They obtained the best-fit parameters  k = $1380^{+134}_{-121}$ and m = $0.47 \pm 0.12$. The corresponding linear fit is shown as the dashed (green) line in their figure \ref{m-k}. Using this relation together with the fluence values of Zone-A derived from our Fermi-GBM analysis, we estimated the pseudo-redshifts for the six sGRBs in our sample. These pseudo-redshifts, along with the associated luminosity distances, are listed in Table \ref{pseudo-tab} (marked with asterisks). The table also includes the two sGRBs with measured redshifts for comparison.

We estimate the event rate of sGRBs that resemble sGRB 170817A using their luminosity distances. The rate is calculated as, $ \frac{N} { t \mathbb{D} (4 \pi/3) d_{L}^{3}} Gpc^{-3} yr^{-1} $. where 
t is the time span from 17 July 2008 to 17 August 2017. For simplicity, and due to the uncertainties in the redshift estimates, we take the number of events N to be one for each luminosity distance $d_L$. Figure \ref{event-rate} shows this rate with circles (blue color) representing the sGRBs with measured redshift (z) and with the triangles (red color) indicate the sGRBs with calculated pseudo-redshift ($z_{p}$). 
 
 The event rate clearly fits with a linear line to $d_{L}$ as expected, shown with solid line in figure \ref{event-rate}. We show the fitted relation as a solid green line, obtained using an ordinary least squares (OLS) fit, which yields a slope of 
-3.12 $\pm$ 0.27, and an intercept of -3.42 $\pm$ 0.43. Similarly, using the 11 sGRBs \citep{Liu:2025ezt} in our dataset with measured redshifts \footnote{\url{https://www.mpe.mpg.de/~jcg/grbgen.html}}, we computed their event rates as a function of luminosity distance and fitted the same functional form to this subset. The resulting slope and intercept are -3.01 $\pm$ 0.11 and -3.14 $\pm$ 0.37, respectively.

Using the OLS function fitted above, we have constructed the formula for the possible number of multimessenger (GW+sGRB) events through LIGO and Fermi-GBM. The number of event are calculated for each observational run of, LIGO following, 
\begin{equation}
N_{GW+sGRB}=4 \pi T \int_{d_{Lmin}}^{d_{Lmax}} d_{L}^2 f(d_{L})\  d(d_{L}),
\end{equation}
where T is the total operation time multiplied with the duty cycle of each run. The details of the duration of the run and the duty cycle are given in \citep{2019PhRvX...9c1040A, 2023PhRvX..13d1039A, 2024PhRvD.109b2001A, 2025arXiv250818080T}. $d_{Lmax}$, $d_{Lmin}$ are upper and lower bounds of luminosity distances aimed at each run. We considered $d_{Lmin}$ as 10 Mpc and $d_{Lmax}$ as the maximum BNS observability limit for the respective observation run. We calculated the expected number of GW+sGRB events, for compact binary coalescence (CBC) mergers followed by EM counterparts corresponding to each observation run time, $O_{1}$ to $O_{4}$ for the LIGO-Virgo-KAGRA (LVK) interferometers. The details of the distance range for each run is listed in table \ref{GW+sGRB}. Details of the observability distance for the CBCs are explained in \citep{2020LRR....23....3A, di2025status}. The expected number of GW+sGRB events from the above calculation are listed in the table \ref{GW+sGRB}.

\begin{table}
\fontsize{6}{6}
	\centering
	\caption{Expected number of GW+sGRB events from the LIGO all runs, operation time, and distance calculation, duty cycles are taken from \citep{2020LRR....23....3A, di2025status,2025PhRvD.111f2002C}}
	\label{GW+sGRB}
	\begin{tabular}{ccccc} 
		\hline
		 \textbf{Run} & \textbf{ Time} &\textbf{Duty Cycle}& \textbf{Distance} \textbf{$(d_{L,max})$} & \textbf{Expected number of} \\
         &\textbf{(Days)}&(\%)&\textbf{(Mpc)}&\textbf{GW+sGRB events} \\
		\hline
		$O_1$&80.04&64.6&80&0.58 $\pm 0.05$\\
         \hline
         $O_2$ &171.19&64.6&100& 1.36 $\pm 0.17$ \\
         \hline
         $O_{3a}$ &127.09&71 &110-130&1.07 $\pm 0.23$ \\
         \hline
         $O_{3b}$ &115.34& 79&110-130&0.97 $\pm 0.12$ \\
         \hline
         $O_{4a}$ &169.452&71.2&160-190&1.64 $\pm 0.72$ \\
         \hline
         $O_{4b}$ &389.46&71.2&160-190&3.77 $\pm 1.46$ \\
		\hline
	\end{tabular}
\end{table}

\section{Conclusion and Outlook}\label{discussion}
With the first observation of the GW+sGRB event, sGRB 170817A the origin of sGRB as NS-NS CBC is confirmed. The large inclination, of $\sim 33^{\circ}$ of the GRB jet resulted in a unique feature in the lightcurve, apart from the softer emission. Considering the soft double peak structure of light curve we searched for sGRBs similar to sGRBs in the Fermi-GBM catalog, with a significant statistical approach. Our search yielded around 8 sGRBs including sGRB 150101B. resembling sGRB 170817A. This class of alike sGRBs could constitute the GW+sGRB events for sGRBs originated through NS-NS mergers. Using a pseudo-redshift method we calculated the possible number of events of GW+sGRB for sGRBs through NS-NS origin for different runs of the LIGO. Within $O_{1}$ and $O_{2}$ run the expected such events is $0.58\pm 0.05$ and $1.36 \pm 0.17$ matches very well with the fact that sGRB 170817A was observed during $O_{2}$ run. At the same time, the expected events in $O_{3}$ and $O_{4}$ run altogether give around 6 to 8 events. The possible CBC events reported through NS-NS and NS-BH mergers are GW190425\citep{GW190425_paper}, $GW190917\_114630$\citep{2024PhRvD.109b2001A} and $GW200115\_042309$ \citep{GW200115_paper}. The upcoming $O_5$ run is expected to begin in late 2027 and last for nearly two years. Taking into account the duty cycle of $73.73\%$ (which is the average of the duty cycles of previous two observation runs). To estimate the expected duration of the $O_{5}$
observing run, we applied the linear extrapolation method using the observed time duration of the previous all four completed runs ($O_1$ to $O_4$), we obtained an estimated $O_5$ run time $\sim 884$ days. Using these values, we calculated the expected number of $N_{GW+sGRB}$ events for the distance range of 240 - 330 Mpc to be $7.49 \pm 3.71$.
\section{Acknowledgements}

We thank Soebur Razzaque for helpful discussions. Sanjeeva Rao would like to thank Neha Yadav (M.Sc. student).

\section{Data Availability}

All the sGRBs data used in this work were collected from the NASA HEASARC's Fermi-GBM burst catalog website, following link:  \href{https://heasarc.gsfc.nasa.gov/w3browse/fermi/fermigbrst.html}{Fermi website} . In support of this work, no additional data were generated or analyzed.

\balance
\bibliographystyle{mnras}
\bibliography{Ref}

\begin{thebibliography}{}
\makeatletter
\relax
\def\mn@urlcharsother{\let\do\@makeother \do\$\do\&\do\#\do\^\do\_\do\%\do\~}
\def\mn@doi{\begingroup\mn@urlcharsother \@ifnextchar [ {\mn@doi@} {\mn@doi@[]}}
\def\mn@doi@[#1]#2{\def\@tempa{#1}\ifx\@tempa\@empty \href {http://dx.doi.org/#2} {doi:#2}\else \href {http://dx.doi.org/#2} {#1}\fi \endgroup}
\def\mn@eprint#1#2{\mn@eprint@#1:#2::\@nil}
\def\mn@eprint@arXiv#1{\href {http://arxiv.org/abs/#1} {{\tt arXiv:#1}}}
\def\mn@eprint@dblp#1{\href {http://dblp.uni-trier.de/rec/bibtex/#1.xml} {dblp:#1}}
\def\mn@eprint@#1:#2:#3:#4\@nil{\def\@tempa {#1}\def\@tempb {#2}\def\@tempc {#3}\ifx \@tempc \@empty \let \@tempc \@tempb \let \@tempb \@tempa \fi \ifx \@tempb \@empty \def\@tempb {arXiv}\fi \@ifundefined {mn@eprint@\@tempb}{\@tempb:\@tempc}{\expandafter \expandafter \csname mn@eprint@\@tempb\endcsname \expandafter{\@tempc}}}

\bibitem[\protect\citeauthoryear{Abbott et~al.}{Abbott et~al.}{2017}]{LIGOScientific:2017vwq}
Abbott B.~P.,  et~al., 2017, \mn@doi [Phys. Rev. Lett.] {10.1103/PhysRevLett.119.161101}, 119, 161101

\bibitem[\protect\citeauthoryear{{Abbott} et~al.,}{{Abbott} et~al.}{2019}]{2019PhRvX...9c1040A}
{Abbott} B.~P.,  et~al., 2019, \mn@doi [Physical Review X] {10.1103/PhysRevX.9.031040}, \href {https://ui.adsabs.harvard.edu/abs/2019PhRvX...9c1040A} {9, 031040}

\bibitem[\protect\citeauthoryear{{Abbott} et~al.,}{{Abbott} et~al.}{2020a}]{2020LRR....23....3A}
{Abbott} B.~P.,  et~al., 2020a, \mn@doi [Living Reviews in Relativity] {10.1007/s41114-020-00026-9}, \href {https://ui.adsabs.harvard.edu/abs/2020LRR....23....3A} {23, 3}

\bibitem[\protect\citeauthoryear{Abbott et~al.}{Abbott et~al.}{2020b}]{GW190425_paper}
Abbott B.~P.,  et~al., 2020b, \mn@doi [Astrophys. J. Lett.] {10.3847/2041-8213/ab75f5}, 892, L3

\bibitem[\protect\citeauthoryear{Abbott et~al.}{Abbott et~al.}{2021}]{GW200115_paper}
Abbott R.,  et~al., 2021, \mn@doi [Astrophys. J. Lett.] {10.3847/2041-8213/ac082e}, 915, L5

\bibitem[\protect\citeauthoryear{Abbott et~al.}{Abbott et~al.}{2023a}]{KAGRA:2021duu}
Abbott R.,  et~al., 2023a, \mn@doi [Phys. Rev. X] {10.1103/PhysRevX.13.011048}, 13, 011048

\bibitem[\protect\citeauthoryear{{Abbott} et~al.,}{{Abbott} et~al.}{2023b}]{2023PhRvX..13d1039A}
{Abbott} R.,  et~al., 2023b, \mn@doi [Physical Review X] {10.1103/PhysRevX.13.041039}, \href {https://ui.adsabs.harvard.edu/abs/2023PhRvX..13d1039A} {13, 041039}

\bibitem[\protect\citeauthoryear{{Abbott} et~al.,}{{Abbott} et~al.}{2024}]{2024PhRvD.109b2001A}
{Abbott} R.,  et~al., 2024, \mn@doi [\prd] {10.1103/PhysRevD.109.022001}, \href {https://ui.adsabs.harvard.edu/abs/2024PhRvD.109b2001A} {109, 022001}

\bibitem[\protect\citeauthoryear{{Amati} et~al.,}{{Amati} et~al.}{2002}]{2002A&A...390...81A}
{Amati} L.,  et~al., 2002, \mn@doi [\aap] {10.1051/0004-6361:20020722}, \href {https://ui.adsabs.harvard.edu/abs/2002A&A...390...81A} {390, 81}

\bibitem[\protect\citeauthoryear{Antunes, Ribeiro, Gomes  \& Aguiar}{Antunes et~al.}{2018}]{antunes2018knee}
Antunes M.,  Ribeiro J.,  Gomes D.,   Aguiar R.~L.,  2018, in 2018 IEEE 6th international conference on future internet of things and cloud (FiCloud). pp 413--419

\bibitem[\protect\citeauthoryear{{Atteia}}{{Atteia}}{2003}]{2003A&A...407L...1A}
{Atteia} J.-L.,  2003, \mn@doi [\aap] {10.1051/0004-6361:20030958}, \href {https://ui.adsabs.harvard.edu/abs/2003A&A...407L...1A} {407, L1}

\bibitem[\protect\citeauthoryear{{Belczynski}, {Taam}, {Kalogera}, {Rasio}  \& {Bulik}}{{Belczynski} et~al.}{2007}]{2007ApJ...662..504B}
{Belczynski} K.,  {Taam} R.~E.,  {Kalogera} V.,  {Rasio} F.~A.,   {Bulik} T.,  2007, \mn@doi [\apj] {10.1086/513562}, \href {https://ui.adsabs.harvard.edu/abs/2007ApJ...662..504B} {662, 504}

\bibitem[\protect\citeauthoryear{Bhat et~al.}{Bhat et~al.}{2016}]{Bhat:2016odd}
Bhat P.~N.,  et~al., 2016, \mn@doi [Astrophys. J. Suppl.] {10.3847/0067-0049/223/2/28}, 223, 28

\bibitem[\protect\citeauthoryear{Burns, Veres, Mészáros, Connaughton, Briggs, Goldstein  et~al.}{Burns et~al.}{2018a}]{Burns:2018qwx}
Burns E.,  Veres P.,  Mészáros P.,  Connaughton V.,  Briggs M.~S.,  Goldstein A.,   et~al., 2018a, \mn@doi [Astrophys. J. Lett.] {10.3847/2041-8213/aad813}, 863, L34

\bibitem[\protect\citeauthoryear{Burns et~al.,}{Burns et~al.}{2018b}]{burns2018fermi}
Burns E.,  et~al., 2018b, The Astrophysical Journal Letters, 863, L34

\bibitem[\protect\citeauthoryear{{Capote} et~al.,}{{Capote} et~al.}{2025}]{2025PhRvD.111f2002C}
{Capote} E.,  et~al., 2025, \mn@doi [\prd] {10.1103/PhysRevD.111.062002}, \href {https://ui.adsabs.harvard.edu/abs/2025PhRvD.111f2002C} {111, 062002}

\bibitem[\protect\citeauthoryear{{Cash}}{{Cash}}{1979}]{1979ApJ...228..939C}
{Cash} W.,  1979, \mn@doi [\apj] {10.1086/156922}, \href {https://ui.adsabs.harvard.edu/abs/1979ApJ...228..939C} {228, 939}

\bibitem[\protect\citeauthoryear{Coulter et~al.}{Coulter et~al.}{2017}]{Coulter:2017wya}
Coulter D.~A.,  et~al., 2017, \mn@doi [Science] {10.1126/science.aap9811}, 358, 1556

\bibitem[\protect\citeauthoryear{{Coward} et~al.,}{{Coward} et~al.}{2012}]{2012MNRAS.425.2668C}
{Coward} D.~M.,  et~al., 2012, \mn@doi [\mnras] {10.1111/j.1365-2966.2012.21604.x}, \href {https://ui.adsabs.harvard.edu/abs/2012MNRAS.425.2668C} {425, 2668}

\bibitem[\protect\citeauthoryear{Cui et~al.}{Cui et~al.}{2020}]{cui2020introduction}
Cui M.,  et~al., 2020, Accounting, Auditing and Finance, 1, 5

\bibitem[\protect\citeauthoryear{De~Maesschalck, Jouan-Rimbaud  \& Massart}{De~Maesschalck et~al.}{2000}]{de2000mahalanobis}
De~Maesschalck R.,  Jouan-Rimbaud D.,   Massart D.~L.,  2000, Chemometrics and intelligent laboratory systems, 50, 1

\bibitem[\protect\citeauthoryear{Del~Vecchio, Dainotti  \& Ostrowski}{Del~Vecchio et~al.}{2016a}]{del2016study}
Del~Vecchio R.,  Dainotti M.~G.,   Ostrowski M.,  2016a, The Astrophysical Journal, 828, 36

\bibitem[\protect\citeauthoryear{Del~Vecchio, Dainotti  \& Ostrowski}{Del~Vecchio et~al.}{2016b}]{DelVecchio:2016vjn}
Del~Vecchio R.,  Dainotti M.~G.,   Ostrowski M.,  2016b, \mn@doi [Astrophys. J.] {10.3847/0004-637X/828/1/36}, 828, 36

\bibitem[\protect\citeauthoryear{Di~Cesare}{Di~Cesare}{2025}]{di2025status}
Di~Cesare M.,  2025, arXiv preprint arXiv:2505.18802

\bibitem[\protect\citeauthoryear{Drout et~al.}{Drout et~al.}{2017}]{Drout:2017ijr}
Drout M.~R.,  et~al., 2017, \mn@doi [Science] {10.1126/science.aaq0049}, 358, 1570

\bibitem[\protect\citeauthoryear{Eichler, Livio, Piran  \& Schramm}{Eichler et~al.}{1989}]{Eichler:1989ve}
Eichler D.,  Livio M.,  Piran T.,   Schramm D.~N.,  1989, \mn@doi [Nature] {10.1038/340126a0}, 340, 126

\bibitem[\protect\citeauthoryear{Evans et~al.}{Evans et~al.}{2017}]{Evans:2017mmy}
Evans P.~A.,  et~al., 2017, \mn@doi [Science] {10.1126/science.aap9580}, 358, 1565

\bibitem[\protect\citeauthoryear{Fok \& Ye}{Fok \& Ye}{2024}]{fok2024deep}
Fok T.~Y.,  Ye N.,  2024, arXiv preprint arXiv:2409.15608

\bibitem[\protect\citeauthoryear{Fong et~al.}{Fong et~al.}{2017}]{Fong:2017ekk}
Fong W.,  et~al., 2017, \mn@doi [Astrophys. J. Lett.] {10.3847/2041-8213/aa9018}, 848, L23

\bibitem[\protect\citeauthoryear{{Goldstein} et~al.,}{{Goldstein} et~al.}{2017}]{2017ApJ...848L..14G}
{Goldstein} A.,  et~al., 2017, \mn@doi [\apjl] {10.3847/2041-8213/aa8f41}, \href {https://ui.adsabs.harvard.edu/abs/2017ApJ...848L..14G} {848, L14}

\bibitem[\protect\citeauthoryear{Granot, Gill, Guetta  \& De~Colle}{Granot et~al.}{2018}]{Granot:2017gwa}
Granot J.,  Gill R.,  Guetta D.,   De~Colle F.,  2018, \mn@doi [Mon. Not. Roy. Astron. Soc.] {10.1093/mnras/sty2308}, 481, 1597

\bibitem[\protect\citeauthoryear{Gruber et~al.}{Gruber et~al.}{2014}]{Gruber:2014iza}
Gruber D.,  et~al., 2014, \mn@doi [Astrophys. J. Suppl.] {10.1088/0067-0049/211/1/12}, 211, 12

\bibitem[\protect\citeauthoryear{Hayes, Heng, Lamb, Lin, Veitch  \& Williams}{Hayes et~al.}{2023}]{Hayes:2023zxm}
Hayes F.,  Heng I.~S.,  Lamb G.,  Lin E.-T.,  Veitch J.,   Williams M.~J.,  2023, \mn@doi [Astrophys. J.] {10.3847/1538-4357/ace899}, 954, 92

\bibitem[\protect\citeauthoryear{Howell, Ackley, Rowlinson  \& Coward}{Howell et~al.}{2019}]{Howell:2018nhu}
Howell E.~J.,  Ackley K.,  Rowlinson A.,   Coward D.,  2019, \mn@doi [Mon. Not. Roy. Astron. Soc.] {10.1093/mnras/stz455}, 485, 1435

\bibitem[\protect\citeauthoryear{{Jiang} et~al.,}{{Jiang} et~al.}{2025}]{2025NatCo..16.2668J}
{Jiang} L.-Y.,  et~al., 2025, \mn@doi [Nature Communications] {10.1038/s41467-025-57791-w}, \href {https://ui.adsabs.harvard.edu/abs/2025NatCo..16.2668J} {16, 2668}

\bibitem[\protect\citeauthoryear{Jin \& Han}{Jin \& Han}{2011}]{jin2011k}
Jin X.,  Han J.,  2011, Encyclopedia of machine learning, pp 563--564

\bibitem[\protect\citeauthoryear{Kalogera, Narayan, Spergel  \& Taylor}{Kalogera et~al.}{2001}]{Kalogera:2001dz}
Kalogera V.,  Narayan R.,  Spergel D.~N.,   Taylor J.~H.,  2001, \mn@doi [Astrophys. J.] {10.1086/321583}, 556, 340

\bibitem[\protect\citeauthoryear{Kalogera et~al.}{Kalogera et~al.}{2004}]{Kalogera:2003tn}
Kalogera V.,  et~al., 2004, \mn@doi [Astrophys. J. Lett.] {10.1086/425868}, 601, L179

\bibitem[\protect\citeauthoryear{Kaneko, Preece, Briggs, Paciesas, Meegan  \& Band}{Kaneko et~al.}{2006}]{Kaneko:2006ru}
Kaneko Y.,  Preece R.~D.,  Briggs M.~S.,  Paciesas W.~S.,  Meegan C.~A.,   Band D.~L.,  2006, \mn@doi [AIP Conf. Proc.] {10.1063/1.2207873}, 836, 133

\bibitem[\protect\citeauthoryear{Kapadia, Dimple, Jain, Misra, Arun  \& Lekshmi}{Kapadia et~al.}{2024}]{Kapadia:2024elj}
Kapadia S.~J.,  Dimple Jain D.,  Misra K.,  Arun K.~G.,   Lekshmi R.,  2024, \mn@doi [Astrophys. J. Lett.] {10.3847/2041-8213/ad8dc7}, 976, L10

\bibitem[\protect\citeauthoryear{Komatsu et~al.}{Komatsu et~al.}{2009}]{WMAP:2008lyn}
Komatsu E.,  et~al., 2009, \mn@doi [Astrophys. J. Suppl.] {10.1088/0067-0049/180/2/330}, 180, 330

\bibitem[\protect\citeauthoryear{{Kumar} \& {Sharma}}{{Kumar} \& {Sharma}}{2024}]{2024arXiv241113242K}
{Kumar} A.,  {Sharma} K.,  2024, \mn@doi [arXiv e-prints] {10.48550/arXiv.2411.13242}, \href {https://ui.adsabs.harvard.edu/abs/2024arXiv241113242K} {p. arXiv:2411.13242}

\bibitem[\protect\citeauthoryear{Li \& Wu}{Li \& Wu}{2012}]{li2012clustering}
Li Y.,  Wu H.,  2012, Physics Procedia, 25, 1104

\bibitem[\protect\citeauthoryear{Liu, Zhang, Dong, Li  \& Du}{Liu et~al.}{2025}]{Liu:2025ezt}
Liu Y.,  Zhang Z.-B.,  Dong X.-F.,  Li L.-B.,   Du X.-Y.,  2025

\bibitem[\protect\citeauthoryear{Lloyd}{Lloyd}{1982}]{lloyd1982least}
Lloyd S.,  1982, IEEE transactions on information theory, 28, 129

\bibitem[\protect\citeauthoryear{MacQueen}{MacQueen}{1967}]{macqueen1967some}
MacQueen J.,  1967, in Proceedings of the Fifth Berkeley Symposium on Mathematical Statistics and Probability, Volume 1: Statistics. pp 281--298

\bibitem[\protect\citeauthoryear{Mahalanobis}{Mahalanobis}{2018}]{mahalanobis2018generalized}
Mahalanobis P.~C.,  2018, Sankhy{\=a}: The Indian Journal of Statistics, Series A (2008-), 80, S1

\bibitem[\protect\citeauthoryear{Mallozzi, Preece  \& Briggs}{Mallozzi et~al.}{2005}]{mallozzi2005rmfit}
Mallozzi R.,  Preece R.,   Briggs M.,  2005, Univ. Alabama, Huntsville

\bibitem[\protect\citeauthoryear{{Margalit} \& {Metzger}}{{Margalit} \& {Metzger}}{2017}]{2017ApJ...850L..19M}
{Margalit} B.,  {Metzger} B.~D.,  2017, \mn@doi [\apjl] {10.3847/2041-8213/aa991c}, \href {https://ui.adsabs.harvard.edu/abs/2017ApJ...850L..19M} {850, L19}

\bibitem[\protect\citeauthoryear{{Margutti} et~al.,}{{Margutti} et~al.}{2017}]{2017ApJ...848L..20M}
{Margutti} R.,  et~al., 2017, \mn@doi [\apjl] {10.3847/2041-8213/aa9057}, \href {https://ui.adsabs.harvard.edu/abs/2017ApJ...848L..20M} {848, L20}

\bibitem[\protect\citeauthoryear{Mazwi, Razzaque  \& Nyadzani}{Mazwi et~al.}{2024}]{Mazwi:2024ego}
Mazwi L.,  Razzaque S.,   Nyadzani L.,  2024, \mn@doi [Mon. Not. Roy. Astron. Soc.] {10.1093/mnras/stae1312}, 531, 2162

\bibitem[\protect\citeauthoryear{Meegan, Koshut  \& Preece}{Meegan et~al.}{1998}]{Meegan:1998tn}
Meegan C.~A.,  Koshut T.~M.,   Preece R.~D.,  eds, 1998, {Proceedings, 4th Huntsville Symposium: Gamma-Ray Bursts}: {Huntsville, USA, September 15-20, 1997}  Vol. 428

\bibitem[\protect\citeauthoryear{Mooley et~al.,}{Mooley et~al.}{2018}]{Mooley:2018qfh}
Mooley K.~P.,  et~al., 2018, \mn@doi [Nature] {10.1038/s41586-018-0486-3}, 561, 355

\bibitem[\protect\citeauthoryear{Na, Xumin  \& Yong}{Na et~al.}{2010}]{na2010research}
Na S.,  Xumin L.,   Yong G.,  2010, in 2010 Third International Symposium on intelligent information technology and security informatics. pp 63--67

\bibitem[\protect\citeauthoryear{Narayan, Piran  \& Shemi}{Narayan et~al.}{1991}]{Narayan:1991fn}
Narayan R.,  Piran T.,   Shemi A.,  1991, \mn@doi [Astrophys. J. Lett.] {10.1086/186143}, 379, L17

\bibitem[\protect\citeauthoryear{Narayan, Paczynski  \& Piran}{Narayan et~al.}{1992}]{Narayan:1992iy}
Narayan R.,  Paczynski B.,   Piran T.,  1992, \mn@doi [Astrophys. J. Lett.] {10.1086/186493}, 395, L83

\bibitem[\protect\citeauthoryear{Pandey, Gupta, Chandra  \& Sathyaprakash}{Pandey et~al.}{2025}]{Pandey:2024mlo}
Pandey S.,  Gupta I.,  Chandra K.,   Sathyaprakash B.~S.,  2025, \mn@doi [Astrophys. J. Lett.] {10.3847/2041-8213/add15f}, 985, L17

\bibitem[\protect\citeauthoryear{Pian et~al.}{Pian et~al.}{2017}]{Pian:2017gtc}
Pian E.,  et~al., 2017, \mn@doi [Nature] {10.1038/nature24298}, 551, 67

\bibitem[\protect\citeauthoryear{{Pizzichini}, {Ferrero}, {Genghini}, {Gianotti}  \& {Topinka}}{{Pizzichini} et~al.}{2006}]{2006AdSpR..38.1338P}
{Pizzichini} G.,  {Ferrero} P.,  {Genghini} M.,  {Gianotti} F.,   {Topinka} M.,  2006, \mn@doi [Advances in Space Research] {10.1016/j.asr.2005.11.016}, \href {https://ui.adsabs.harvard.edu/abs/2006AdSpR..38.1338P} {38, 1338}

\bibitem[\protect\citeauthoryear{Poggiani et~al.}{Poggiani et~al.}{2020}]{poggiani2020multi}
Poggiani R.,  et~al., 2020, POS PROCEEDINGS OF SCIENCE, 362

\bibitem[\protect\citeauthoryear{Poolakkil et~al.}{Poolakkil et~al.}{2021}]{Poolakkil:2021jpc}
Poolakkil S.,  et~al., 2021, \mn@doi [Astrophys. J.] {10.3847/1538-4357/abf24d}, 913, 60

\bibitem[\protect\citeauthoryear{Porciani \& Madau}{Porciani \& Madau}{2001}]{Porciani:2000ag}
Porciani C.,  Madau P.,  2001, \mn@doi [Astrophys. J.] {10.1086/319027}, 548, 522

\bibitem[\protect\citeauthoryear{Qumsiyeh \& Sabha}{Qumsiyeh \& Sabha}{2023}]{qumsiyeh2023utilizing}
Qumsiyeh E.,  Sabha M.,  2023, in 2023 2nd International Engineering Conference on Electrical, Energy, and Artificial Intelligence (EICEEAI). pp~1--7

\bibitem[\protect\citeauthoryear{Salafia, Ghisellini, Pescalli, Ghirlanda  \& Nappo}{Salafia et~al.}{2016}]{Salafia:2016wru}
Salafia O.~S.,  Ghisellini G.,  Pescalli A.,  Ghirlanda G.,   Nappo F.,  2016, \mn@doi [Mon. Not. Roy. Astron. Soc.] {10.1093/mnras/stw1549}, 461, 3607

\bibitem[\protect\citeauthoryear{Satopaa, Albrecht, Irwin  \& Raghavan}{Satopaa et~al.}{2011}]{satopaa2011finding}
Satopaa V.,  Albrecht J.,  Irwin D.,   Raghavan B.,  2011, in 2011 31st international conference on distributed computing systems workshops. pp 166--171

\bibitem[\protect\citeauthoryear{Schubert}{Schubert}{2023}]{schubert2023stop}
Schubert E.,  2023, ACM SIGKDD Explorations Newsletter, 25, 36

\bibitem[\protect\citeauthoryear{Smartt et~al.}{Smartt et~al.}{2017}]{Smartt:2017fuw}
Smartt S.~J.,  et~al., 2017, \mn@doi [Nature] {10.1038/nature24303}, 551, 75

\bibitem[\protect\citeauthoryear{Syakur, Khotimah, Rochman  \& Satoto}{Syakur et~al.}{2018}]{syakur2018integration}
Syakur M.~A.,  Khotimah B.~K.,  Rochman E.,   Satoto B.~D.,  2018, in IOP conference series: materials science and engineering. p. 012017

\bibitem[\protect\citeauthoryear{{Tanvir} et~al.,}{{Tanvir} et~al.}{2017}]{2017ApJ...848L..27T}
{Tanvir} N.~R.,  et~al., 2017, \mn@doi [\apjl] {10.3847/2041-8213/aa90b6}, \href {https://ui.adsabs.harvard.edu/abs/2017ApJ...848L..27T} {848, L27}

\bibitem[\protect\citeauthoryear{{The LIGO Scientific Collaboration} et~al.,}{{The LIGO Scientific Collaboration} et~al.}{2025}]{2025arXiv250818080T}
{The LIGO Scientific Collaboration} et~al., 2025, \mn@doi [arXiv e-prints] {10.48550/arXiv.2508.18080}, \href {https://ui.adsabs.harvard.edu/abs/2025arXiv250818080T} {p. arXiv:2508.18080}

\bibitem[\protect\citeauthoryear{Troja et~al.}{Troja et~al.}{2017}]{Troja:2017nqp}
Troja E.,  et~al., 2017, \mn@doi [Nature] {10.1038/nature24290}, 551, 71

\bibitem[\protect\citeauthoryear{Umargono, Suseno  \& Gunawan}{Umargono et~al.}{2020}]{umargono2020k}
Umargono E.,  Suseno J.~E.,   Gunawan S.~V.,  2020, in The 2nd international seminar on science and technology (ISSTEC 2019). pp 121--129

\bibitem[\protect\citeauthoryear{{Villar} et~al.,}{{Villar} et~al.}{2017}]{2017ApJ...851L..21V}
{Villar} V.~A.,  et~al., 2017, \mn@doi [\apjl] {10.3847/2041-8213/aa9c84}, \href {https://ui.adsabs.harvard.edu/abs/2017ApJ...851L..21V} {851, L21}

\bibitem[\protect\citeauthoryear{Von~Kienlin et~al.,}{Von~Kienlin et~al.}{2019}]{von2019fermi}
Von~Kienlin A.,  et~al., 2019, The Astrophysical Journal, 876, 89

\bibitem[\protect\citeauthoryear{Zhang et~al.}{Zhang et~al.}{2018}]{Zhang:2017lpb}
Zhang B.~B.,  et~al., 2018, \mn@doi [Nature Commun.] {10.1038/s41467-018-02847-3}, 9, 447

\bibitem[\protect\citeauthoryear{{Zitouni}, {Guessoum}  \& {Azzam}}{{Zitouni} et~al.}{2022}]{2022Ap&SS.367...74Z}
{Zitouni} H.,  {Guessoum} N.,   {Azzam} W.,  2022, \mn@doi [\apss] {10.1007/s10509-022-04100-2}, \href {https://ui.adsabs.harvard.edu/abs/2022Ap&SS.367...74Z} {367, 74}

\bibitem[\protect\citeauthoryear{{von Kienlin} et~al.,}{{von Kienlin} et~al.}{2019}]{2019ApJ...876...89V}
{von Kienlin} A.,  et~al., 2019, \mn@doi [\apj] {10.3847/1538-4357/ab10d8}, \href {https://ui.adsabs.harvard.edu/abs/2019ApJ...876...89V} {876, 89}

\makeatother
\end{thebibliography}

\onecolumn
\begin{table}
\fontsize{6}{6}
	\centering
	\caption{Pseudo-redshift, luminosity distance, $E_{peak}^{\text{i}}$, $E_{iso}$, and  Energy fluence ($E_{f}$) for the sGRBs without redshift.{\textsuperscript{\ding{72}}}The sGRBs with known redshift.}
	\label{pseudo-tab}
	\begin{tabular}{cccccccc} 
		\hline
		\textbf{Name} & \textbf{$z$ or $z_{p}$} & \textbf{$d_{L} $} &\textbf{$E_{peak}^{\text{i}}$} & \textbf{$E_{iso}$}&\textbf{$E_{f} (Zone-A)$}&\textbf{$E_{f} (Zone-B)$}& \textbf{Chances of occurrence}\\
&&$(Gpc)$&$(keV)$&$(\mathrm{erg})\times 10^{48}$&$(erg /cm^{2}) \times 10^{-7}$&$(erg /cm^{2})\times 10^{-7}$&$(10^{-2})$\\
		\hline
		GRB 170817529{\textsuperscript{\ding{72}}} & $0.009783 \pm 0.0071$ & $0.0422 \pm 0.018$ & $178.02 \pm 78.87$  & $0.059 \pm 0.003$  &$1.20 \pm 0.40 $&$0.266 \pm 0.02 $&0.28\\
GRB 150805746 & $0.18 \pm 0.03$ & $0.87 \pm 0.17$ & $90.10 \pm 12.50$ & $31.75 \pm 10.24$ &$ 3.40\pm 0.37 $&$ 1.10\pm 0.19 $&0.25\\
GRB 150101641{\textsuperscript{\ding{72}}} & $0.134 \pm 0.058 $ & $0.6330 \pm 0.454$ &  $184.61 \pm 10.70$ & $10.06 \pm 0.63$ &$ 0.26\pm 0.18 $&$ 0.21\pm 0.07 $&3.29\\
GRB 131128629 & $0.15 \pm 0.04 $ & $0.75 \pm 0.24$ & $91.79 \pm 22.24$ & $35.70 \pm 20.99$ &$ 2.20\pm 0.23 $&$ 0.82\pm 0.15 $&3.12\\
GRB 131004904{\textsuperscript{\ding{72}}} & $0.717 \pm  0.130  $ & $4.4529 \pm 0.789$& $184.23 \pm 76.31$ & $704.83 \pm 26.39$&$ 4.66\pm 0.67 $& $ 0.86\pm 0.13  $&10.48\\
GRB 130808253 & $0.27 \pm 0.05$& $1.41 \pm 0.35$ &$105.80 \pm 18.66$  & $46.41 \pm 22.03$  &$3.10 \pm0.72 $&$ 1.1\pm0.21  $&0.28\\
GRB 120524134 & $0.18 \pm 0.04 $ & $0.88 \pm 0.23$ & $75.20 \pm 14.11$ & $21.73 \pm 10.36$ &$2.69 \pm 0.28 $&$ 0.37\pm 0.07 $&0.55\\
GRB 090108020 & $0.32 \pm 0.04$& $1.71 \pm 0.28$& $185.83 \pm 20.05$ &  $143.60 \pm 41.01$&$0.78\pm 0.12 $&$ 0.74 \pm  0.07$&3.12\\
GRB 081122614 & $0.93 \pm 0.16$ & $7.97 \pm 1.71$& $277.60 \pm 35.55$ & $1132.90 \pm 246.20$ &$0.61 \pm0.19  $&$ 0.04\pm 0.004 $&3.40\\
		\hline
	\end{tabular}
\end{table}
\begin{longtable}{cccccccc}
\caption{\small Hardness ratio's, $HR_{1}$ and $HR_{2}$ of all the sGRBs listed in our final set of data. Associated times ($T_{90}$,$T_{50}$) and detectors that we considered for analysis. Along with the redshift value in the last column. \\
    NOTE: * We did not consider the GRB for K-means clustering.} \\
\hline
 \textbf{GRBs} & \textbf{$T_{90}^{\text{start}}$ (s)} & \textbf{$T_{50}^{\text{end}}$ (s)} & \textbf{$T_{90}^{\text{end}}$ (s)} & \textbf{Detectors} & \textbf{$HR_1$} & \textbf{$HR_2$} & \textbf{Redshift value($z^c$)}  \\
\hline
\endfirsthead
\multicolumn{8}{c}{{\textit{Table \thetable\ (continued)}}} \\
\hline
 \textbf{GRBs} & \textbf{$T_{90}^{\text{start}}$ (s)} & \textbf{$T_{50}^{\text{end}}$ (s)} & \textbf{$T_{90}^{\text{end}}$ (s)} & \textbf{Detectors} & \textbf{$HR_1$} & \textbf{$HR_2$} & \textbf{Redshift value($z^c$)}  \\
\hline
\endhead
\hline \multicolumn{8}{c}{r}{{Continued on next page...}} \\

\endfoot


\endlastfoot

GRB170817529 & -0.192 & 1.216 & 1.856 & n2, n5 & 0.578 & 0.0784&0.009783h \\
\hline
GRB170726249 & -0.768 & 0.512 & 1.024 & n7, n9 & 1.508 & 1.0911&- \\
GRB170511648 & -1.28 & -0.512 & 0 & n6,n11 & 1.587 & 2.280& -\\
GRB170304003 &-0.016  &0.048  &0.144  & n1, n9 & 2.222 &0.6029 & -\\
GRB161026373 & -0.032 & 0.032 & 0.080 & n0, n3 & 1.440 & 0.682& -\\
GRB160826938 & -1.344 & -0.192 & 0.448 & n1, n2 & 0.809 &0.534 & -\\
GRB160821937 & -0.064 & 0.32 & 1.024 & n9,n11 & 0.0616 & 1.024&0.16h \\
GRB160603719 & -0.128 & 0.128 & 0.256 & n7, n8 & 3.617 & 1.038 &-\\
GRB160411062 & -0.928 & -0.384 & -0.256 & n4, n8 & 1.598 &0.557 & -\\
GRB160224911 & -0.064 & 0.320 & 0.320 & n1, n5 & 13.529 & 0.7897& -\\
GRB151229486 & -0.064 & 0.032 & 0.096 & n3, n4 & 3.401 & 0.624 &1.4ph\\
GRB150923864 & 0 & 1.088 & 1.792 & n7, n8 & 1.120 & 1.2847& -\\
GRB150923429 & -0.064 & 0.064 & 0.128 & n4, n9 & 12.471 & 1.102& -\\

GRB150912600 & -0.128 & 0.192 & 0.192 & n1, n2 & 11.934 & 0.379& -\\

GRB150901924 &  -0.064& 0.064 & 0.192 & n10, n11 &  3.384& 2.965 & -\\
GRB150805746 & -0.576 & 0.192 & 0.832 & n7, n11 & 0.534 & 0.187& -\\
GRB150705588 & -0.256 & 0 & 0.448 & n3, n7 & 1.121 & 0.595& -\\
GRB150628767 & -0.064 & 0.256 & 0.576 & n6, n8 & 5.039 & 2.916 &-\\

GRB150609316 & -0.240 & -0.016 & 0.016 & n0, n1 & 0.145 & 1.979&- \\
GRB150605782 & -0.048 & -0.016 & 0.128 & n7, n8 & 1.142 & 0.780& -\\
GRB150412507 & -0.128 & 0.128 & 0.448 & n8, n9 &1.0813  &3.485 & -\\
GRB150325696 & -0.032 & 0.016 & 0.048 & n6, n7 & 7.770  &3.035 &- \\

GRB150301045 &  -0.032&  0.032& 0.384 & n4, n8 & 1.095 & 1.094& -\\
GRB150101270 & -0.416 & -0.032 & 0.064 & n0,n3 & 0.589 & 3.82& - \\
GRB150101641 &-0.016  & 0 & 0.064 & n8, n9 & 1.805 & 0.126& 0.134h\\
GRB141128962 & -0.096 & 0.032 & 0.172 & n2, n5 & 1.503 &0.161 & -\\
GRB141102112 &-0.032 & -0.016&  -0.016&  n6,n8&2.918 & 0.810 & -\\ 
GRB140724533& -0.256 & 0.064 & 0.64 & n8, n11 & 1.345 & 1.757 &-\\
GRB140624423 &  -0.080&  0.016&  0.016& n3, n4 & 2.148 &1.601 & -\\
GRB140209313 & 1.344 & 1.856 & 2.752 & n9, n10 & 1.4822 & 0.518& -\\
GRB131128629 & -0.96 & 0.256 & 1.024 & n10,n11 & 0.869 & 0.308& -\\


GRB131004904
 & -0.192 &  0.384&  0.960& n9, n10 & 0.7065 & 0.635&0.717  \\
GRB130919173 & -0.064 & 0.768 & 0.896 & n7, n8 & 0.947 & 0.722& -\\
GRB130808253 & -0.128 & 0 & 0.128 & n4,n5 & 0.710 & 0.284&- \\
GRB130622615 & -0.768 & 0 & 0.192 & n7, n8 & 1.108 & 1.031 &-\\
GRB130617564 & -0.448 & 0.064 & 0.32 & n2, n10 & 1.835 & 1.283 &-\\
GRB130404877 & -0.128 & 0.192 & 0.832 & n0, n1 & 1.639 & 0.225&- \\
GRB130325005 & -0.064 & 0.192 & 0.576 & n6,n7 & 3.676 & 1.575&- \\

GRB121004211 & -0.512 & 0.512 &  1.024& n3, n4 & 0.847 & 1.226& -\\

GRB120831901 & -0.256 & 0 & 0.128 & n4, n5 &1.855 &2.280 &-\\
GRB120814803$^*$  &  -0.192&0  &0  & n4, n6 & 28.87 & 1.112& -\\
GRB120629565 & -0.384 & 0.064 & 0.32 & n3, n6 & 5.040 & 0.373& -\\
GRB120524134 & -0.128 & 0.128 & 0.576 & n4, n8 & 0.546 & 0.188 -\\
GRB120429003 & -0.192 & 0.640 & 1.472 & n10, n11 & 1.757 &1.385 & -\\
GRB120327418 & -0.192 & -0.064 & 0.064 & n4, n8 & 0.063 & 3.076 &2.813\\
GRB120222021 & -0.064 & 0.576 & 1.024 & n3,n5 & 1.123 & 0.789&- \\
GRB120101354 & -0.096 & 0 & 0.032 & n7, n8 & 1.701 & 0.053 & -\\
GRB111207512 & -0.896 & -0.256 & -0.128 & n0, n1 & 8.557 & 0.715 &-\\

GRB111117510 & -0.064 & 0.352 & 0.368 & n0, n9 & 2.328 & 2.062 &2.21h\\
GRB111011094 & -0.064 & 0.128 & 1.408 & n8, n11 & 1.457 & 0.541 &-\\
GRB111001804 & -0.256 & 0 & 0.128 & n7, n11 & 9.018 & 2.031 &-\\

GRB110605780 & -0.256 & 0.768 & 1.28 & n8, n10 & 3.056 & 0.669& -\\
GRB110509475 & -0.320 & -0.064 & 0.320 & n3, n5 & 1.448 & 0.556 &-\\
GRB110227009 &-0.192&0.640  &1.536 & n7, n8 &  3.577 & 0.246 &  -\\
GRB110131780 & -0.192 & 0.064 & 0.192 & n0, n1 & 7.219 & 1.887&- \\
GRB101216721 & 0.003 & 0.832 & 1.92 & n1, n5 & 1.490 & 0.0628& -\\
GRB101208498 &  -0.640&  0.640& 1.408 & n2, n5 & 0.387 &0.694 & -\\

GRB101027230 & -1.280 & 0.064 & 0.064 & n6, n7 & 3.892 & 0.657&- \\
GRB100816026 & 0.003 & 1.408 & 2.048 & n8,n11 & 1.67 & 1.311&0.8035 \\
GRB100805300 & -0.096 & -0.032 & -0.032 & n8, n10 & 2.061 & 2.315& -\\

GRB100629801 & -0.128 & 0.320 & 0.704 & n10, n11 & 1.259 & 0.618& -\\
GRB100616773 & -0.192 & -0.064 & 0 & n9, n10 & 6.260 & 1.163&- \\
GRB100516396 & -0.576 & -0.064 & 0.064 & n6,n7 & 3.374 & 2.490& -\\
GRB100417166 & -0.064 & 0.064 & 0.128 & n7,n9 & 2.364 & 0.770&- \\
GRB100117879 & -0.064 & 0.064 & 0.192 & n4, n8 & 2.938 & 2.278&0.92phh \\
GRB091126333 & -0.064 & 0.064 & 0.128 & n6, n11 & 2.860 & 2.200& -\\
GRB091019750 &-0.112 &-0.016  & 0.096 &n1,n2 &2.004 & 1.544 & -\\
GRB091006360 & -0.192 & -0.064 & 0 & n1, n2 & 2.592 &1.223 & -\\

GRB090927422 & -0.192 & 0.192 & 0.32 & n2, n10 &1.938  & 1.503& 1.37\\
GRB090616157 & -0.192 & 0.512 & 0.960 & n2, n5 & 0.893 & 0.465& -\\
GRB090518080 & -0.640 & 0.768 & 1.408 & n1, n3 & 0.1124 &1.516 & -\\

GRB090108020 & -0.064 & 0.256 & 0.64 & n1,n2 & 1.038 & 0.324&- \\

GRB081230871 & -0.128 & 0.192 & 0.384 & n7, n9 & 7.577 & 2.049& 2ph\\

GRB081226044 & -0.192 & 0.192 & 0.64 & n2, n10 & 1.837 & 4.007& -\\
GRB081223419 & -0.064 & 0.256 & 0.512 & n7, n11 & 1.424 & 0.8896&- \\
GRB081122614 &-0.064 & 0.064&  0.128 & n4, n7 &0.634 & 0.702& -\\
GRB081107321 & -0.192 & 1.152 & 1.472 & n6,n9 & 0.577 & 0.585& -\\
GRB081101491$^*$  & -0.064 & 0.064 & 0.064 & n9, n10 & 2.294 & 7.249&- \\












GRB080919790 & -0.128 & 0.064 & 0.384 & n0,n2 & 2.075 & 0.671& -\\
GRB080831053 & -0.288 & 0 & 0.288 & n0, n5 & 4.616 & 0.789& - \\ \hline

\label{table-hr}
\end{longtable}

\begin{longtable}{ccccccc}
             \caption{ Best fit parameters for all the 80 sGRBs from the final sample set.
From left to right, the \textbf{first column} contains the name of GRB; in the \textbf{second column}, the first time interval is for the Zone-A that was observed in the GRB, which is fitted with the Comp model, and corresponding parameter values $E_{peak}$, $\alpha$ are listed in the \textbf{fourth and fifth columns}, respectively. The second time interval is for the Zone-B that's seen in the GRB, which is fitted with the BB model, and the corresponding best fitted parameter values (kT) are given in the \textbf{sixth column}. In the \textbf{seventh column}, C-stat and degree of freedom (dof) values are noted to describe how well the model fits the data.} \\
\hline
    \textbf{GRB Names} & \textbf{Time (s)} & \textbf{Model} & \textbf{E$_{\text{peak}}$ (keV)} & \textbf{Index} & \textbf{kT (keV)} & \textbf{C-stat/dof}  \\
    \hline
    \endfirsthead

    \multicolumn{7}{c}{{\textit{Table \thetable\ (continued)}}} \\
    \hline
    \textbf{GRB Names} & \textbf{Time (s)} & \textbf{Model} & \textbf{Epeak (keV)} & \textbf{Index} & \textbf{kT (keV)} & \textbf{C-stat/dof}  \\
    \hline
    \endhead

    \hline \multicolumn{7}{r}{{Continued on next page...}} \\  
    \endfoot


\endlastfoot

 \textbf{GRB170817529} & -0.192:1.216 & Comp & 176.3 $\pm$ 85.7 & -0.9364 $\pm$ 0.168 & -    &   247.55/231\\
& -0.192:1.216 & PL & -  & $-1.586 \pm 0.162 $& -  &   248.64/232\\
              &  1.216:1.856& BB   &-&-&9.236 $\pm$ 1.83 &263.06/232\\
              &  1.216:1.856& PL   &-&- 2.140$\pm$ 0.482&- &269.93/232\\
\hline
\hline
GRB170726249 & -0.768:0.512&Comp & 481.4 $\pm$ 398& -0.933 $\pm$ 0.32& -  &  210.31/227 \\
             & 0.512:1.024  & BB  &   - & - &34.64 $\pm$ 0.26& 243.3/228 \\
\hline
GRB170511648 & -0.08:0.016 & Comp &44.9 $\pm$ 54.9 & 0.2861 $\pm$ 1.5& - & 263.67/231 \\
             & 0.016:0.016& BB & -  & - & 56.03 $\pm$ 15& 270.84/232 \\ 
             \hline
              GRB170304003 & -0.064:-0.016 & Comp & 272.2 $\pm$ 149& 0.043 $\pm$ 0.005& -    &  196.81/229\\
              &  -0.016:0.032 & BB   &-&-&5.233 $\pm$ 0.03& 162.10/230\\
\hline
GRB161026373& -0.0.32:0.032& Comp &324.6 $\pm$ 181 & -0.503 $\pm$ 0.44& - &229.1/228  \\
             & 0.032:0.080& BB & - & - & 49.71 $\pm$ 7.95& 269.6/229 \\
\hline
\textbf{GRB160826938} & -1.344:-0.192 & Comp &  99.90 $\pm$ 22.2& -0.3554 $\pm$ 0.49&  -& 244.39/228 \\
             & -0.192:0.448& BB &-  &-  & 14.23 $\pm$ 2.63&  271.67/229\\
\hline
\textbf{GRB160821937} & -0.064:0.32& Comp  & 79.44 $\pm$ 18.2&-0.9368 $\pm$ 0.38& - &267.04/228  \\
             &0.32:1.024 & BB & - & - & 29.64 $\pm$ 8.18&253.01/229  \\
\hline
GRB160603719 & -0.128-0.128 & Comp & 377.2 $\pm$ 151& 0.191 $\pm$ 0.629& -    &   296.9/228 \\
              &  0.128-0.256& BB   &-&-&21.35 $\pm$ 7.86&232.87/229\\
\hline
GRB160411062 & -0.384:-0.256 & Comp & 83.76 $\pm$ 15.9& -0.9362 $\pm$ 0.31&  -  &   226.83/227 \\
&1.344-1.664&BB&-&-&9.550 $\pm$ 1.15& 254.14/228\\
\hline
GRB160224911 & -0.064:0.32& Comp  & 308.9$\pm$ 188&  0.045$\pm$ 0.012& - & 204.92/221 \\
             & 0.32:0.32& BB & - & - & 46.34 $\pm$ 59.1 &235.99/222  \\        
             \hline
             GRB150923864 & 0:1.088 &Comp & 131.7 $\pm$ 10.7& -0.243 $\pm$ 0.15& -  &  227.40/229 \\
             & 1.088:1.792  & BB  &   - & - &25.89 $\pm$ 1.91& 240.84/230 \\
\hline
GRB150923429 & -0.064:0.064& Comp  & 229.8$\pm$103 &  0.0784$\pm$ 2.04& - &217.55 / 221\\
             & 0.064:0.128& BB & - & - &  49.21$\pm$15.3  &182.54/222  \\        
             \hline
GRB150912600 & -0.128:0.192& Comp  &660.2 $\pm$304.7 &-0.463  $\pm$0.274 & - & 281.74/241 \\
             &0.192 :0.192& BB & - & - &  37.80$\pm$4.79  &282.67/242  \\        
             \hline             
GRB150901924 & -0.08:0.016 & Comp &289.8 $\pm$ 93.4 & 1.412 $\pm$ 1.15& - & 224.46/225 \\
             & 0.016:0.016& BB & -  & - &21.44 $\pm$ 6.50& 202.61/226 \\
\hline

\textbf{GRB150805746\textsuperscript{\ding{72}}} & -0.576:0.192 & Comp & 76.35 $\pm$14.7 & 1.226 $\pm$ 0.28 & -   &  251.34/228  \\
& 0.192:0.832 & BB  & -  &-  &9.201 $\pm$ 4.72&248.76/229   \\
& 0.192:0.832&PL&-& $-2.61 \pm 0.30$&- & 257.83/229\\
\hline
GRB150705588 & -0.256:0&  Comp& 152.8 $\pm$ 66.6& -0.9619 $\pm$ 0.38& - & 219.12/226 \\
             & 0:0.448 &  BB&  -& - &15.18 $\pm$ 2.99& 232.97/227 \\
             
\hline
GRB150628767 & 0.034:0.022& Comp  & 267.7$\pm$75.8 & 0.630 $\pm$ 0.812& - & 218.93/223 \\
             & 0.256:0.576& BB & - & - &  62.74$\pm$ 13.6 &267.33/224  \\        
             \hline 
\textbf{GRB150609316} & -0.24:-0.016 & Comp & 5457 $\pm$ 1677& -1.034 $\pm$ 0.786 & - &231.36/228\\
             & -0.016-0.016 & BB  & - & -  & 39.04 $\pm$ 5.46&231.71/229\\
\hline
GRB150605782 & -0.048:-0.016& Comp  & 251.72$\pm$ 80.25& -0.369$\pm$ 0.06& - &243.34/227  \\
             &-0.016:0.128 & BB & - & - & 44.47 $\pm$ 7.30&218.64/228  \\
             \hline
GRB150412507 & -0.128:0.128 & Comp &89.75 $\pm$ 28& -0.9054 $\pm$ 0.54& - & 237.84/223 \\
             & 0.128:0.448& BB & -  & - & 16.08 $\pm$ 6.23& 223.61/ 224\\
\hline
GRB150325696 & -0.0.32:0.016& Comp & 248.5 $\pm$ 63.5&  0.599 $\pm$ 0.65& - & 263.42/229 \\
             &0.016:0.048 &  BB& - &-  &43.89 $\pm$ 6.03& 232.16/230 \\
\hline
GRB150301045 & -0.032:0.032 & Comp & 257.6 $\pm$ 128& -0.7496 $\pm$ 0.41& -   &  230.20/231  \\
& 0.032:0.384 & BB  & -  &-  &14.76 $\pm$ 4.14&228.16/232   \\

\hline
 GRB150101270 & -0.416:-0.032& Comp  & 224.6 $\pm$ 89.4&0.2787 $\pm$ 0.08& - &221.12/232  \\
             &-0.032:0.064 & BB & - & - &  45.31 $\pm$ 9.06&241.44/233 \\
\hline
\textbf{GRB150101641\textsuperscript{\ding{72}}}& -0.016:0 & Comp & 162.8 $\pm$ 18.3& -0.9619 $\pm$ 0.29& - &241.80/230\\
             & 0:0.064& BB  & - & -  & 9.828 $\pm$ 4.15&198.71/231\\
             &0:0.064&PL&-&-2.588&-&203.26/231\\

\hline
GRB141128962 & -0.096:0.032& Comp  & 167.6$\pm$48.8 &  -0.489$\pm$0.381 & - & 204.76/217 \\
             & 0.032:0.176& BB & - & - &  13.06$\pm$1.17  &258.22/218  \\        
             \hline 
GRB141102112 & -0.032:-0.016 & Comp & 87.46 $\pm$ 115& 0.197 $\pm$0.52 & -    &   305.82/253\\
              &  -0.016:-0.016& BB   &-&-&9.822 $\pm$ 1.62&280.88/254\\
\hline
GRB140724533 &-0.256:0.064& Comp  & 207.3$\pm$114 &  -0.696$\pm$0.480 & - & 262.05/230 \\
             & 0.064:0.64& BB & - & - &  29.05$\pm$12.2  &277.82/231  \\        
             \hline
GRB140624423 & -0.08:0.016 & Comp &260.6  $\pm$ 95.6& -0.7626 $\pm$ 0.30& - & 241.4/228 \\
             & 0.016:0.016& BB & -  & -  2.84& 33.80$\pm$ 2.84& 223.54/221 \\
\hline
GRB140209313 & 1.344:1.856 & Comp & 189.1 $\pm$ 51.8 &  -0.262 $\pm$ 0.04&  -&292.8/227  \\
             & 1.856:2.752 & BB & - & - & 14.88 $\pm$ 0.195& 217/228 \\
\hline
\textbf{GRB131128629\textsuperscript{\ding{72}}} & -0.96:0.256 & Comp & 79.41 $\pm$ 51.33& 0.086 $\pm$0.13 & - &216.92/225\\
             & 0.256:1.024 & BB  & - & -  & 10.41 $\pm$ 0.97&291.17/226\\
             &0.256:1.024&PL&-&-1.981&-&295.81/226\\
\hline
\textbf{GRB131004904\textsuperscript{\ding{72}}} & -0.192:0.384& Comp &107.3 $\pm$ 30.1 & -1.197 $\pm$ 0.26& - &230.44/225  \\
             & 0.384:0.96& BB & - & - & 7.761 $\pm$ 0.75& 250.22/226 \\
             &0.256:1.024&PL&-&$-2.31 \pm 0.20$&-&252.85/226\\

\hline
\textbf{GRB130919173} & -0.064:0.768  &Comp  & 141.5 $\pm$ 28.6 & -0.5225 $\pm$ 0.313&-  &236.60/220\\
             & 0.768:0.896 &BB  & - & -  & 18.04 $\pm$ 1.63&216.34/221\\
\hline
\textbf{GRB130808253\textsuperscript{\ding{72}}} & -0.128:0 & Comp & 83.05 $\pm$ 41.28 &  -0.11 $\pm$ 0.05 &  -&244.23/224  \\
             & 0:0.128 & BB & - & - & 12.00 $\pm$ 5.17 & 201.81/225 \\
             &0:0.128 &PL&-&$-1.991 \pm 0.16$&-&210.69/225\\
\hline
GRB130622615 & -0.768-0 &Comp  & 134 $\pm$ 33.5& -0.531 $\pm$ 0.39&-  &254.4/229\\
             & 0-0.192 &BB  & - & -  & 31.89 $\pm$ 5.45 &230.9/230\\
\hline
GRB130617564 & -0.448:0.064& Comp  & 1127$\pm$ 1000&  -0.772$\pm$ 0.294& - & 216.39/223 \\
             & 0.064:0.320& BB & - & - &  41.29$\pm$15.9  &257.94/224  \\        
             \hline 
             
GRB130404877 & -0.128:0.192& Comp &229.6 $\pm$ 93.8& -0.032 $\pm$ 0.073& - & 232.1/226 \\
             &0.192:0.832 & BB & - & - &73.24 $\pm$ 32.7&  215.45/227\\
\hline
GRB130325005 & -0.064:0.192& Comp  &231.5 $\pm$ 87.4 &1.98 $\pm$1.67 & - & 222.15/230 \\
             &0.192 :0.576& BB & - & - &  47.21$\pm$16.13   &226.22/231  \\        
             \hline
\textbf{GRB121004211}& -0.512:0.512 & Comp &  125.8 $\pm$ 29.9& -0.5616 $\pm$ 0.35&-  &246.6/224\\
             & 0.512:1.024 & BB & - & -  &19.62 $\pm$ 2.21 &245.27/230\\
\hline 
GRB120831901 & 502.1:431& Comp  & 502.1 $\pm$431.0 &-0.330$\pm$0.690   & - & 255.28/226 \\
             & 0:0.128& BB & - & - &  49.89$\pm$7.05  &208.82/227  \\        
             \hline
GRB120814803 & -0.192:0& Comp  &170.2 $\pm$ 38.4& 2.647 $\pm$ 2.81& - &243.34/227  \\
             &0:0 & BB & - & - & 64.88$\pm$ 21.56& 203.03/228  \\
\hline 
GRB120629565 & -0.384:0.064& Comp  &200.8 $\pm$89 & 0.325 $\pm$ 0.157& - &254.84/229  \\
             &0.064:0.320 & BB & - & - & 18.6 $\pm$ 9.13&246.3/230  \\
\hline
\textbf{GRB120524134\textsuperscript{\ding{72}}} & -0.128:0.128 & Comp & 63.65 $\pm$ 7.88& -0.2814 $\pm$ 0.46 & -&259.40/228 \\
             & 0.128:0.576& BB & - & -  &7.6 $\pm$ 0.85&258.60/229\\    &0.128:0.576 &PL&-&$-2.28 \pm 0.29$&-&262.26/229\\         
\hline

GRB120429003 & -0.192:0.64 & Comp & 576.6 $\pm$ 503&  -1.106 $\pm$ 0.215&  -&292.8/227  \\
             & 0.64:1.472 & BB & - & - & 14.16 $\pm$ 1.87& 224.80/228 \\
\hline
GRB120327418 &-0.192 :-0.064& Comp  & 309.7$\pm$268 &  -0.055$\pm$1.31 & - &260.64 /223 \\
             & -0.064:0.064& BB & - & - &  38.45$\pm$7.99  &256.72/224  \\        
             \hline 
GRB120222021 & -0.064:0.576 &Comp  & 139.1 $\pm$ 95.62& -0.501 $\pm$ 0.64  &-  &243.02/253\\
             & 0.576:1.024 &BB  & - & -  & 18.41 $\pm$ 11.88 &324.71/254\\
\hline
GRB120101354 & -0.096:0& Comp  & 185.65$\pm$ 65.4&  0.180$\pm$0.742 & - &214.57 /222 \\
             & 0:0.032& BB & - & - & 27.62 $\pm$ 7.06 &176.82/223  \\        
             \hline
GRB111207512 & -0.896:-0.256& Comp  & 730.11$\pm$817.50 &  -0.0781$\pm$ 1.07& - & 258.04/227 \\
             & -0.256:-0.128& BB & - & - & 1.624 $\pm$0.43  &222.98/228  \\        
             \hline
 GRB111117510 &-0.064 :0.352& Comp  &519.6 $\pm$222 &  -0.492$\pm$0 & - & 280.64/224 \\
             & 0.352:0.368& BB & - & - &  71.91$\pm$9.67  & 210.20/225  \\        
             \hline   
GRB111011094 & -0.064:0.128& Comp  &286.8 $\pm$76.2 & -0.379 $\pm$0.261 & - & 278.21/225 \\
             & 0.128:1.408& BB & - & - &  11.53$\pm$6.31  &257.58/226  \\        
             \hline    
            
GRB111001804 & -0.256:0& Comp  & 504.3$\pm$490 &  -0.332$\pm$0.868 & - & 256.90/223 \\
             & 0:0.128& BB & - & - &  30.09$\pm$  8.88&205.96/224  \\        
             \hline 
 GRB110605780 &-0.256 :0.768& Comp  & 181.7$\pm$57.6 &  -0.105$\pm$0.69 & - & 256.98/225 \\
             & 0.768:1.28& BB & - & - &  20.223$\pm$6.99  &270.30/226  \\        
             \hline
 GRB110509475 & -0.32:0.32& Comp  & 403.8$\pm$284 &  -0.910$\pm$ 0.339& - & 236.66/233 \\
             & 0.32:0.32& BB & - & - &  15.37$\pm$3.45  &255.28/ 224 \\        
             \hline
             GRB110227009 & -0.192:0.640 & Comp  & 99.72 $\pm$ 9.65& 0.202 $\pm$ 0.36&-  &216.43/221\\
             & 0.640:1.53 & BB &-  &-   &11.05 $\pm$ 1.23&231.45/222\\
             
\hline
GRB110131780 & -0.192:0.064& Comp  &239.1 $\pm$72.7 & 1.88 $\pm$1.97 & - & 239.12/216 \\
             & 0.064:0.192& BB & - & - &  12.66$\pm$1.55  &215.15/217  \\        
             \hline 
              
GRB101216721 & 0.003:0.832& Comp  &108.5 $\pm$ 29.64& -0.348 $\pm$ 0.17& - &272.27/253  \\
             &0.832:1.92 & BB & - & - & 13.26 $\pm$6.15 &394.2/254  \\        \hline
 \textbf{GRB101208498} & -0.64:0.64 & Comp & 133.2 $\pm$ 9.08& -0.941 $\pm$ 0.067& -    &   234.07/226 \\
              &  0.64:1.408 & BB   &-&-&7.847 $\pm$ 2.07&211.67/227\\
\hline

GRB101027230 & -1.28:0.064& Comp &409.8 $\pm$ 121.6& 1.78 $\pm$ 0.73& - & 241.4/228 \\
             & 0.064:0.064& BB & -  & - & 29.14$\pm$ 7.12& 224.6/229 \\
    \hline
 GRB100816026 & 0.003:1.408 & Comp &  146.7 $\pm$79.52 & -0.244 $\pm$0.19 &-  &253.11/253\\
             & 1.408:2.048 & BB & - & -  &24.71 $\pm$ 9.1&365.22/254\\
\hline   
GRB100805300&-0.096 :-0.032& Comp  & 189$\pm$77.4 &  1.412$\pm$5.6 & - & 142.99/128 \\
             & -0.032:-0.032& BB & - & - &  52.62$\pm$1.2  &131.61/127  \\        
             \hline 
 GRB100629801 & -0.128-0.32 & Comp &  183.61 $\pm$ 25.7 & -0.581 $\pm$ 0.198 &-  &246.6/224\\
             & 0.32-0.704 & BB & - & -  &31.14 $\pm$ 4 &267.1/225\\
\hline  
GRB100616773 & -0.192:-0.064& Comp  &472.8 $\pm$306 & -0.341 $\pm$ 0.619& - & 249.38/222 \\
             & -0.064:0& BB & - & - &  54.59$\pm$8.87  &231.62/223  \\        
             \hline 
GRB100516396 & -0.576:-0.064&Comp &200.1 $\pm$ 121.79  & 9.399 $\pm$0.597 & -  &  278.46/253\\
             & -0.064:0.064 &  BB& - & -  & 54.72 $\pm$ 29.13&232.6/254\\
\hline
GRB100417166 & -0.048:-0.016& Comp  &224.4 $\pm$ 60.8& -0.5763 $\pm$ 0.571& - &206.55/226  \\
             &-0.016:0.128 & BB & - & - & 48.11 $\pm$ 18.1&197.80/227\\
\hline
GRB100117879 & -0.064:0.064& Comp  & $328.6\pm$132 &  -0.2533$\pm$0.0 & - & 231.74/ 206\\
             &0.064:0.192& BB & - & - & 35.19 $\pm$6.11  &226.37/227  \\        
             \hline
GRB091126333 & -0.064:0.064& Comp  & $177.1\pm$38.4 & 0.115 $\pm$ 0.586& - & 257.52/ 230\\
             & 0.064:0.128& BB & - & - &  38.81$\pm$3.87  &252.45/ 231 \\        
             \hline    
GRB091019750 & -0.112:-0.016& Comp  & 696$\pm$ 636&  -0.402$\pm$0.609 & - & 243.50/217 \\
             & -0.016:0.096& BB & - & - &  73.32$\pm$19.5  &203.47/218  \\        
             \hline

GRB091019750 & -0.112:-0.016 & Comp & 388.5 $\pm$ 161& -0.6045 $\pm$ 2.33& -   &  264.2/225  \\
& -0.016:0.096 & BB  & -  &-  &16.11 $\pm$ 2.33&228.16/226   \\
\hline
GRB091006360 & -0.192:-0.064 & Comp & 193.1 $\pm$ 79.61 & -0.508 $\pm$ 0.23&  -  &   222.5/253 \\
&-0.064:0&BB&-&-&67.68 $\pm$ 45.63& 247.95/254\\
\hline
GRB090927422 & -0.192:0.192 & Comp & 106.7 $\pm$ 19.2& 0.460 $\pm$ 0.07& -    &  256.15/226\\
              &  0.192:0.32 & BB   &-&-&25.91 $\pm$ 6.15 &209.32/227\\
\hline

\textbf{GRB090616157} & -0.192:0.512& Comp & 1576 $\pm$ 709&  -1.370 $\pm$ 0.27& - & 206.75/229 \\
             &0.512:0.96 &  BB& - &-  &52.98 $\pm$ 35.8& 269.58/230 \\
\hline
\textbf{GRB090518080} & -0.64:768 & Comp &286.0 $\pm$ 163 & -1.235 $\pm$ 0.18& - & 224.53/229 \\
             & 0.768:1.408& BB & -  & - & 28.83 $\pm$ 4.46& 217.93/230 \\
\hline
\textbf{GRB090108020\textsuperscript{\ding{72}}} & -0.064:0.256& Comp &140.3  $\pm$ 87.35& -0.424 $\pm$ 0.19& - &226.05/223  \\
             & 0.256:0.64& BB & - & - & 10.73 $\pm$ 6.93& 242.93/224 \\
             &0.256:0.64 &PL&-&$-1.87 \pm 0.13$&-&238.29/224\\
\hline
GRB081230871 & -0.128:0.192& Comp  & 320.60$\pm$96.1 &  2.050$\pm$ 1.90& - & 234.01/226 \\
             & 0.192:0.384& BB & - & - &  92.35$\pm$ 17.3 &226.28/ 227 \\        \hline 
GRB081226044 & -0.192:0.192& Comp  & 532.7$\pm$417 & -0.959 $\pm$0.271 & - & 207.44/ 222 \\
            & 0.192:0.64& BB & - & - & 38.29 $\pm$ 4.86 &208.99/223\\        
            \hline 
GRB081223419 & -0.064:0.256&  Comp& 197 $\pm$ 30.5& -0.429 $\pm$ 0.21& - & 230/226 \\
             & 0.256:0.512&  BB&  -& - &23.78 $\pm$ 2.6& 259.8/227 \\
\hline

\textbf{GRB081122614\textsuperscript{\ding{72}}} & -0.128:0.192& Comp & 143.8 $\pm$ 22.0 & -1.331 $\pm$ 0.39 & - & 193.67/221 \\
            &0.192:0.832 & BB & - & - &11.26 $\pm$ 6.98 & 193.52 /222\\
            &0.192:0.832 &PL&-&$-1.93 \pm 1.71$&-&194.03/222\\
\hline
\textbf{GRB081107321} & -0.192:1.152 & Comp  & 72.73 $\pm$ 38.19 & -0.3865 $\pm$ 0.097 &-  &317.48/253\\
             & 1.152:1.472 & BB &-  &-   &17.17 $\pm$9.85 &285.96/254\\
\hline
GRB081101491 & -0.064-0.064& Comp & 551.8 $\pm$ 437.1&0.835 $\pm 0.331$ & -  &  238.6/226\\
             & 0.064-0.06 &  BB& - & -  &  37.6 $\pm$ 9.69 & 241.4/227\\
\hline




GRB080919790 & -0.128:0.064& Comp & 183.6 $\pm$116.7 &  -0.2307 $\pm$0.072 & - & 232.48/253 \\
             &0.064:0.384 &  BB& - &-  &37.23 $\pm$ 9.66& 283.91/254 \\
\hline

GRB080831053 & -0.288:0& Comp  & 154$\pm$41.2 &  1.305$\pm$ 1.89& - & 207.31/ 230 \\
             & 0:0.288& BB & - & - & 41.92 $\pm$ 10.4&221.39/231  \\        
             \hline 

\label{table-spec}
\end{longtable}







\bsp	
\label{lastpage}

\end{document}